\documentclass[12pt,a4paper]{article}

\usepackage{ifthen} 
\newboolean{pdflatex}
\setboolean{pdflatex}{true} 

\newboolean{articletitles}
\setboolean{articletitles}{true} 

\newboolean{inbibliography}
\setboolean{inbibliography}{false} 

\usepackage[top=1in, bottom=1.25in, left=1in, right=1in]{geometry}

\usepackage{microtype}
\usepackage{lineno}  
\usepackage{xspace} 
\usepackage{caption} 

\usepackage{graphicx}  
\usepackage{color}
\usepackage{colortbl}
\graphicspath{{./figs/}} 

\usepackage{amsmath} 
\usepackage{amssymb}
\usepackage{amsfonts}
\usepackage{upgreek} 

\newcommand*\patchAmsMathEnvironmentForLineno[1]{%
\expandafter\let\csname old#1\expandafter\endcsname\csname #1\endcsname
\expandafter\let\csname oldend#1\expandafter\endcsname\csname
end#1\endcsname
 \renewenvironment{#1}%
   {\linenomath\csname old#1\endcsname}%
   {\csname oldend#1\endcsname\endlinenomath}%
}
\newcommand*\patchBothAmsMathEnvironmentsForLineno[1]{%
  \patchAmsMathEnvironmentForLineno{#1}%
  \patchAmsMathEnvironmentForLineno{#1*}%
}
\AtBeginDocument{%
\patchBothAmsMathEnvironmentsForLineno{equation}%
\patchBothAmsMathEnvironmentsForLineno{align}%
\patchBothAmsMathEnvironmentsForLineno{flalign}%
\patchBothAmsMathEnvironmentsForLineno{alignat}%
\patchBothAmsMathEnvironmentsForLineno{gather}%
\patchBothAmsMathEnvironmentsForLineno{multline}%
\patchBothAmsMathEnvironmentsForLineno{eqnarray}%
}

\usepackage{hyperref}    
\usepackage[all]{hypcap} 

\usepackage{ifthen} 
\newboolean{uprightparticles}
\setboolean{uprightparticles}{false} 
\usepackage{xspace} 
\usepackage{upgreek}

\def\MagUp {\mbox{\em Mag\kern -0.05em Up}\xspace}

\ifthenelse{\boolean{uprightparticles}}%
{

 \def\Pmu         {\ensuremath{\upmu}\xspace}

 \def\Ppi         {\ensuremath{\uppi}\xspace}

 \def\Ppsi        {\ensuremath{\uppsi}\xspace}

 \def\PDelta      {\ensuremath{\Delta}\xspace}                 
 \def\PXi      {\ensuremath{\Xi}\xspace}                 
 \def\PLambda      {\ensuremath{\Lambda}\xspace}                 
 \def\PSigma      {\ensuremath{\Sigma}\xspace}                 
 \def\POmega      {\ensuremath{\Omega}\xspace}                 
 \def\PUpsilon      {\ensuremath{\Upsilon}\xspace}                 
 
 \def\PB      {\ensuremath{\mathrm{B}}\xspace}                 
                  
 \def\PD      {\ensuremath{\mathrm{D}}\xspace}

 \def\PJ      {\ensuremath{\mathrm{J}}\xspace}                 
 \def\PK      {\ensuremath{\mathrm{K}}\xspace}

 \def\Pb      {\ensuremath{\mathrm{b}}\xspace}                 
 \def\Pc      {\ensuremath{\mathrm{c}}\xspace}                 
                  
 \def\Pe      {\ensuremath{\mathrm{e}}\xspace}

 \def\Pi      {\ensuremath{\mathrm{i}}\xspace}

 \def\Pp      {\ensuremath{\mathrm{p}}\xspace}

 \def\Ps      {\ensuremath{\mathrm{s}}\xspace}

}
{

 \def\Pmu         {\ensuremath{\mu}\xspace}

 \def\Ppi         {\ensuremath{\pi}\xspace}

 \def\Ppsi        {\ensuremath{\psi}\xspace}                 
                  
 \mathchardef\PDelta="7101
 \mathchardef\PXi="7104
 \mathchardef\PLambda="7103
 \mathchardef\PSigma="7106
 \mathchardef\POmega="710A
 \mathchardef\PUpsilon="7107
                  
 \def\PB      {\ensuremath{B}\xspace}                 
                  
 \def\PD      {\ensuremath{D}\xspace}

 \def\PJ      {\ensuremath{J}\xspace}                 
 \def\PK      {\ensuremath{K}\xspace}

 \def\Pb      {\ensuremath{b}\xspace}                 
 \def\Pc      {\ensuremath{c}\xspace}                 
                  
 \def\Pe      {\ensuremath{e}\xspace}

 \def\Pi      {\ensuremath{i}\xspace}

 \def\Pp      {\ensuremath{p}\xspace}

 \def\Ps      {\ensuremath{s}\xspace}

}

\makeatletter
\ifcase \@ptsize \relax
  \newcommand{\miniscule}{\@setfontsize\miniscule{4}{5}}
\or
  \newcommand{\miniscule}{\@setfontsize\miniscule{5}{6}}
\or
  \newcommand{\miniscule}{\@setfontsize\miniscule{5}{6}}
\fi
\makeatother

\DeclareRobustCommand{\optbar}[1]{\shortstack{{\miniscule (\rule[.5ex]{1.25em}{.18mm})}
  \\ [-.7ex] $#1$}}

\def\en         {{\ensuremath{\Pe^-}}\xspace}   
\def\ep         {{\ensuremath{\Pe^+}}\xspace}

\def\mup        {{\ensuremath{\Pmu^+}}\xspace}
\def\mumu       {{\ensuremath{\Pmu^+\Pmu^-}}\xspace}

\def\squark    {{\ensuremath{\Ps}}\xspace}

\def\cquark    {{\ensuremath{\Pc}}\xspace}

\def\bquark    {{\ensuremath{\Pb}}\xspace}

\def\pion   {{\ensuremath{\Ppi}}\xspace}

\def\pip    {{\ensuremath{\pion^+}}\xspace}
\def\pim    {{\ensuremath{\pion^-}}\xspace}

\def\kaon    {{\ensuremath{\PK}}\xspace}
  \def\Kbar    {{\kern 0.2em\overline{\kern -0.2em \PK}{}}\xspace}

\def\KorKbar    {\kern 0.18em\optbar{\kern -0.18em K}{}\xspace}

\def\Kp      {{\ensuremath{\kaon^+}}\xspace}
\def\Km      {{\ensuremath{\kaon^-}}\xspace}

\def\KS      {{\ensuremath{\kaon^0_{\mathrm{ \scriptscriptstyle S}}}}\xspace}

  \def\Dbar    {{\kern 0.2em\overline{\kern -0.2em \PD}{}}\xspace}
\def\D       {{\ensuremath{\PD}}\xspace}

\def\DorDbar    {\kern 0.18em\optbar{\kern -0.18em D}{}\xspace}
\def\Dz      {{\ensuremath{\D^0}}\xspace}
\def\Dzb     {{\ensuremath{\Dbar{}^0}}\xspace}

\def\Dstarz  {{\ensuremath{\D^{*0}}}\xspace}

\def\B       {{\ensuremath{\PB}}\xspace}
\def\Bbar    {{\ensuremath{\kern 0.18em\overline{\kern -0.18em \PB}{}}}\xspace}

\def\BorBbar    {\kern 0.18em\optbar{\kern -0.18em B}{}\xspace}

\def\Bu      {{\ensuremath{\B^+}}\xspace}

\def\Bp      {{\ensuremath{\Bu}}\xspace}

\def\Bs      {{\ensuremath{\B^0_\squark}}\xspace}

\def\jpsi     {{\ensuremath{{\PJ\mskip -3mu/\mskip -2mu\Ppsi\mskip 2mu}}}\xspace}

  \def\Y#1S{\ensuremath{\PUpsilon{(#1S)}}\xspace}

\def\proton      {{\ensuremath{\Pp}}\xspace}

\def\Lbar        {{\ensuremath{\kern 0.1em\overline{\kern -0.1em\PLambda}}}\xspace}
\def\LorLbar    {\kern 0.18em\optbar{\kern -0.18em \PLambda}{}\xspace}

\def\to                 {\ensuremath{\rightarrow}\xspace}

\def\AT#1     {\ensuremath{A_{\mathrm{T}}^{#1}}\xspace}           

\def\C#1      {\ensuremath{\mathcal{C}_{#1}}\xspace}                       
\def\Cp#1     {\ensuremath{\mathcal{C}_{#1}^{'}}\xspace}                    
\def\Ceff#1   {\ensuremath{\mathcal{C}_{#1}^{\mathrm{(eff)}}}\xspace}        
\def\Cpeff#1  {\ensuremath{\mathcal{C}_{#1}^{'\mathrm{(eff)}}}\xspace}       
\def\Ope#1    {\ensuremath{\mathcal{O}_{#1}}\xspace}                       
\def\Opep#1   {\ensuremath{\mathcal{O}_{#1}^{'}}\xspace}                    

\newcommand{\tev}{\ifthenelse{\boolean{inbibliography}}{\ensuremath{~T\kern -0.05em eV}\xspace}{\ensuremath{\mathrm{\,Te\kern -0.1em V}}}\xspace}
\newcommand{\gev}{\ensuremath{\mathrm{\,Ge\kern -0.1em V}}\xspace}
\newcommand{\mev}{\ensuremath{\mathrm{\,Me\kern -0.1em V}}\xspace}
\newcommand{\kev}{\ensuremath{\mathrm{\,ke\kern -0.1em V}}\xspace}
\newcommand{\ev}{\ensuremath{\mathrm{\,e\kern -0.1em V}}\xspace}
\newcommand{\gevc}{\ensuremath{{\mathrm{\,Ge\kern -0.1em V\!/}c}}\xspace}
\newcommand{\mevc}{\ensuremath{{\mathrm{\,Me\kern -0.1em V\!/}c}}\xspace}
\newcommand{\gevcc}{\ensuremath{{\mathrm{\,Ge\kern -0.1em V\!/}c^2}}\xspace}
\newcommand{\gevgevcccc}{\ensuremath{{\mathrm{\,Ge\kern -0.1em V^2\!/}c^4}}\xspace}
\newcommand{\mevcc}{\ensuremath{{\mathrm{\,Me\kern -0.1em V\!/}c^2}}\xspace}

\def\nb {\ensuremath{\mathrm{ \,nb}}\xspace}

\def\gsim{{~\raise.15em\hbox{$>$}\kern-.85em
          \lower.35em\hbox{$\sim$}~}\xspace}
\def\lsim{{~\raise.15em\hbox{$<$}\kern-.85em
          \lower.35em\hbox{$\sim$}~}\xspace}

\def\ptot       {\mbox{$p$}\xspace}
\def\pt         {\mbox{$p_{\mathrm{ T}}$}\xspace}

\def\tell1  {TELL1\xspace}
\def\ukl1   {UKL1\xspace}

\newcommand{\eg}{\mbox{\itshape e.g.}\xspace}

\usepackage{cite} 
\usepackage{mciteplus}

\usepackage{longtable} 
\usepackage{multirow}

\usepackage{cite}
\usepackage{mciteplus}
\usepackage{verbatim}
\begin{document}

\renewcommand{\thefootnote}{\fnsymbol{footnote}}
\setcounter{footnote}{1}


\begin{titlepage}
\pagenumbering{roman}

\vspace*{2.0cm}

{\normalfont\bfseries\boldmath\huge
\begin{center}
  RapidSim: an application for the fast simulation of heavy-quark hadron decays
\end{center}
}

\vspace*{1.5cm}

\begin{center}
G.~A.~Cowan\footnote{g.cowan@ed.ac.uk}, D.~C.~Craik and M.~D.~Needham\\\vspace{1cm}{\it School of Physics and Astronomy \\University of Edinburgh\\ Edinburgh, UK}\\\vspace{1cm}
\today
\end{center}

\vspace{\fill}

\begin{abstract}
  \noindent
  RapidSim is a lightweight application for the fast simulation of phase space decays of 
  beauty and charm quark hadrons, allowing for quick studies of the properties 
  of signal and background decays in particle physics analyses.
  Based upon the {\tt TGenPhaseSpace} class
  from the ROOT application it uses externally provided fixed-order next-to-leading-logarithm
  calculations to boost the initial beauty and charm hadrons 
  to the appropriate energy for the production environment of interest.
  User-defined momentum resolution functions can be used to mimic the
  effect of imperfect track reconstruction. User-defined efficiency shapes can be
  applied during generation to reproduce the effects of geometric and kinematic
  requirements on final state particles as well as the dynamics of the decay.
  The effect of mis-identification of the final 
  state particles is simple to configure via configuration files, while
  the framework can easily be extended to include additional particle types. 
  This paper describes the RapidSim framework, features and some example
  use cases.
\end{abstract}

\vspace*{2.0cm}

\vspace{\fill}

\vspace*{2mm}

\end{titlepage}


\newpage
\setcounter{page}{2}
\mbox{~}


\renewcommand{\thefootnote}{\arabic{footnote}}
\setcounter{footnote}{0}



\pagestyle{plain} 
\setcounter{page}{1}
\pagenumbering{arabic}


%

\section{Introduction}
\label{sec:Introduction}

A common problem in the analysis of particle decays is understanding the
kinematic properties of the
signal decay of interest and the potential backgrounds that can be introduced via other
particle decays that have been imperfectly reconstructed in the detector.
In particular, it is often crucial
to know the shape of the invariant mass of the final state particles (or sub-sets of them)
and the corresponding shape of the background decays. These backgrounds can arise from a number
of sources: 

\begin{itemize}
\item Decays to final states containing a sub-set of the particles in the signal decay;
\item Decays to final states containing the same particles as in the signal decay,
with additional particles that have not been reconstructed by the detector;
\item Decays to final states where some of the particles have been 
misidentified by the detector such that they appear as the signal decay;
\item Some combination of the above.
\end{itemize}

One method to study potential background sources is to generate large samples of the
decays and pass them through the full detector simulation and reconstruction
software chain. By selecting these events as if they are the signal decay,
it is possible to study the shape of the backgrounds in the invariant mass distribution
of interest. The disadvantage of this approach is the typically long time required
to generate, reconstruct and select these background samples and the mass storage
requirements to retain these samples.

RapidSim~\cite{g_a_cowan_2016_160402} is an application that allows analysts to quickly generate large
samples (a few seconds for millions of events) of potential background decays with
momentum spectra,  invariant mass resolutions and efficiency shapes that are close approximations to
what can be obtained from a full detector simulation. The speed of generation
allows analysts to quickly perform initial studies that may indicate avenues for further
investigation that may need a more detailed simulation.
RapidSim supports the generation of a single decay chain with any number of sub-decays.
It utilises the ROOT~\cite{Brun:1997pa} software package and, specifically,
the {\tt TGenPhaseSpace} class to perform the fast generation.
Fixed-order next-to leading-log (FONLL)~\cite{Cacciari:1998it}
calculations are used to provide the correct kinematic properties of the heavy hadrons. 
By default, RapidSim is deployed with appropriate FONLL configuration files such that
heavy hadrons are produced with the kinematic properties that are
expected from proton-proton collisions at the CERN Large Hadron Collider~\cite{Evans:2008zzb}. 
In addition to generating decays in $4\pi$ it is possible for the user to specify that the decays of interest
fall within the geometrical acceptance of the LHCb experiment~\cite{Alves:2008zz}. 
The simple design allows users to extend RapidSim to include alternative detector
geometries.
RapidSim can accommodate basic user-defined kinematic efficiency effects during generation
and mis-identification of final-state particles. 
All of these features are implemented in a generic way to enable them to be configured for any decay chain.

Section~\ref{sec:Usage} of this paper describes how to compile and run the application.
Sections~\ref{sec:Configuration} and~\ref{sec:Bs2Jpsiphi} give an overview of the configuration options and example use case, respectively. Details of the heavy quark production, efficiency effects, track momentum resolution and
particle mis-identification can be found in Sections~\ref{sec:Production},~\ref{sec:Efficiency},~\ref{sec:Smearing} and~\ref{sec:Misid}, respectively. A summary is presented in Section~\ref{sec:Conclusion}.

\section{Usage}
\label{sec:Usage}
The latest version of the source code can be found on GitHub~\cite{g_a_cowan_2016_160402}
and is compiled using the following commands:
\begin{verbatim}
mkdir build
cd build
cmake ..
make
\end{verbatim}

The user should define the environment variable {\tt RAPIDSIM\_ROOT} to point to the root of the {\tt RapidSim}
directory structure. The executable may then be run as
\begin{verbatim}
$RAPIDSIM_ROOT/build/src/RapidSim.exe <decay mode> <events to generate> [<save tree?>]
\end{verbatim}
for example,
\begin{verbatim}
$RAPIDSIM_ROOT/build/src/RapidSim.exe Bs2Jpsiphi 10000 1
\end{verbatim}
where there is a user provided {\tt Bs2Jpsiphi.decay} file and optionally
a user-defined {\tt Bs2Jpsiphi.config} file (described in Sections~\ref{sec:Configuration}
and~\ref{sec:Bs2Jpsiphi}) in the present working directory.

\section{Configuration}
\label{sec:Configuration}

The RapidSim package has been designed to be easily configurable at run time for any decay, with all options relating to the decay configured via a decay-specific configuration file. 
This file is automatically created, with default options, the first occasion a particular decay is generated.
Full lists of the available global and particle-specific settings are given in Tables~\ref{tab:globalsettings} and~\ref{tab:particlesettings}, respectively. For most settings, the syntax is \verb!<setting> : <option>!, where spaces and/or commas may be used to separate multiple options. 
The {\tt param}, {\tt shape} and {\tt cut} global settings require multiple arguments to be provided,
with the following syntax:
\begin{verbatim}
param : <name> <type> <particles> [TRUE]
shape : <file> <hist> <paramX> [<paramY>]
cut : <param> <type> <value1> [<value2>]
\end{verbatim}
For {\tt param}, \verb!<type>! is one of the parameter types defined in Table~\ref{tab:paramtypes}, \verb!<particles>! is a space separated list of particles to combine to calculate the parameter and the optional \verb!TRUE! defines the parameter in terms of true four-momenta rather than reconstructed quantities. 
For {\tt shape}, \verb!<file>! and \verb!<hist>! must give the filesystem location and name
of a {\tt TH1} or {\tt TH2} object, and the \verb!<param{X,Y}>! fields must match the \verb!<name>! field of defined parameters.
For {\tt cut}, \verb!<param>! must match the name of a defined parameter, \verb!<type>! must be one of {\tt min}, {\tt max}, {\tt range} or {\tt veto}, and \verb!<value1>! and \verb!<value2>! define the cutoff values.

\begin{table}[!htb]
\centering
\caption{\small
Global settings configured via the decay configuration file.
}
\label{tab:globalsettings}
\begin{tabular}{lp{0.4\textwidth}p{0.35\textwidth}}
\hline \\ [-2.5ex]
Setting & Description & Options (default) \\
\hline \\ [-2.5ex]
seed & sets the seed for the random number generator & integer values (4357) \\
acceptance & sets the type of geometric acceptance to apply to the decay & (Any), ParentIn, AllIn AllDownstream \\
energy & sets the $pp$ collision energy (in TeV) used to get the correct parent kinematics & 7, (8), 13, 14 \\
parent & overrides the determined flavour of the parent particle used to get the correct parent kinematics & ($b$), $c$ * default is $c$ for decays of charmed hadrons \\
geometry & sets the detector geometry to use for production and acceptance & 4pi, (LHCb) \\
ptRange & sets the range of $\pt$ to generate the parent within & two float values (0 100) \\
etaRange & sets the range of $\eta$ to generate the parent within & two float values (1 6) \\
minWidth & sets the minimum resonance width (in GeV) to be generated. Narrower resonances will be generated with a fixed mass. & float values (0.001) \\
maxAttempts & sets the maximum number of attempts allowed to generate each decay. & integer values (1000) \\
paramsStable & Defines the set of parameters to be added to the histograms/tree for each stable particle in the decay. & (P, PT), M, M2, MT, E, ET, PX, PY, PZ, eta, phi, rapidity, gamma, beta \\
paramsDecaying & Defines the set of parameters to be added to the histograms/tree for each decaying particle in the decay. & (P, PT, M), M2, MT, E, ET, PX, PY, PZ, eta, phi, rapidity, gamma, beta \\
paramsTwoBody & Defines the set of parameters to be added to the histograms/tree for each two-body combination of particles in a 3+-body decay. & P, PT, M, (M2), MT, E, ET, PX, PY, PZ, eta, phi, rapidity, gamma, beta \\
paramsThreeBody & Defines the set of parameters to be added to the histograms/tree for each three-body combination of particles in a 4+-body decay. & P, PT, M, (M2), MT, E, ET, PX, PY, PZ, eta, phi, rapidity, gamma, beta \\
param & Defines a new parameter to be added to the histograms/tree. & \multirow{3}{0.4\textwidth}{See text for details}\\
cut & Applys a cut on a given parameter. & \\
shape & Sets a 1D or 2D PDF to generate events according to. & \\
\hline
\end{tabular}
\end{table}

\begin{table}[!htb]
\centering
\caption{\small
Particle settings configured via the decay configuration file.
}
\label{tab:particlesettings}
\begin{tabular}{lp{0.4\textwidth}p{0.35\textwidth}}
\hline \\ [-2.5ex]
Setting & Description & Options (default) \\
\hline \\ [-2.5ex]
name & A user-friendly name for this particle to be used in variable names. & alphanumeric \\
smear & The type of momentum smearing to apply to this particle. & types of smearing (LHCbGeneric) * defaults to LHCbElectron for $e^\pm$. \\
invisible & Whether the particle should be treated as invisible. Invisible particles are not included when determining non-truth parameters. A corrected mass parameter will be added automatically if any particles are invisible. & true, (false) * defaults to true for neutrinos.\\
altMass & Adds alternative mass hypotheses for this particle. & particle type names \\
\hline
\end{tabular}
\end{table}

\begin{table}[!htb]
\centering
\caption{\small
Types of parameter and their corresponding descriptions.
}
\label{tab:paramtypes}
\begin{tabular}{lp{0.75\textwidth}}
\hline \\ [-2.5ex]
Type & Description \\
\hline \\ [-2.5ex]

M  		    & The invariant mass of the combination of the given particles \\
M2  		& The squared invariant mass of the combination of the given particles \\
MT  		& The transverse mass of the combination of the given particles \\
E  		    & The energy of the combination of the given particles \\
ET  		& The transverse energy of the combination of the given particles \\
P  		    & The total momentum of the combination of the given particles \\
PX  		& The X momentum of the combination of the given particles \\
PY  		& The Y momentum of the combination of the given particles \\
PZ  		& The Z momentum of the combination of the given particles \\
PT  		& The transverse momentum of the combination of the given particles \\
eta  		& The pseudorapidity of the combination \\
phi  		& The azimuthal angle of the combination \\
y         	& The rapidity of the combination \\
gamma  		& The relitivistic gamma factor of the combination \\
beta  		& The velocity of the combination \\
theta  		& The angle between the first two particles in the rest frame of the combination of the remaining particles \eg\ ``1 2 1 3'' would give the angle between 1 and 2 in the rest frame of a resonance in $m_{13}$ \\
costheta  	& The cosine of theta \\
Mcorr  		& The corrected mass of the combination of the given particles correcting for any invisible particles. 
		  Given by $m_{\rm corr} = \sqrt{m^2 + {p_{\rm T}^{\rm miss}}^{2}}+ p_{\rm T}^{\rm miss}$, where $m$ is the reconstructed mass 
		  and $p_{\rm T}^{\rm miss}$ is the momentum of the reconstructed particles transverse to the true flight direction of the combination of all given particles. \\
\hline
\end{tabular}
\end{table}
\clearpage
In addition to the decay-specific configuration file. Several common configuration files are located in the {\tt config/} and {\tt rootfiles/} directories such as
\begin{itemize}
    \item the list of defined particle species;
    \item definitions of production kinematics options;
    \item definitions of momentum resolution options.
\end{itemize}
Further details on the configuration of these latter two features are given in Sections~\ref{sec:Production:fonll} and~\ref{sec:Smearing:config}, respectively.

%


\section{Example: \boldmath $\Bs\to\jpsi\phi$}
\label{sec:Bs2Jpsiphi}

The decay file consists of a single line, which describes the decay using the following syntax:
\begin{itemize}
    \item particles are named according to {\tt config/particles.dat} and separated by spaces;
    \item a decay is denoted by {\tt ->};
    \item subdecays are demarcated by braces {\tt \{\}}.
\end{itemize}
For the  $\Bs\to\jpsi\phi$ decay
the {\tt Bs2Jpsiphi.decay} ASCII file  contains the single line
\begin{verbatim}
Bs0 -> { Jpsi -> mu+ mu- } { phi -> K+ K- }
\end{verbatim}
where the particle names are defined in the {\tt config/particle.dat} file.
In this example the \jpsi ($\phi$) meson decays to two oppositely
charged muons (kaons).
The user-defined configuration file (auto-produced the first time a particular
decay is generated) is written as:
\begin{verbatim}
paramsDecaying : M, P, PT
paramsStable : P, PT

geometry : LHCb

@0
   name : Bs
@1
   name : Jpsi
@2
   name : phi
@3
   name : mup
   smear : LHCbGeneric
@4
   name : mum
   smear : LHCbGeneric
@5
   name : Kp
   smear : LHCbGeneric
   altMass : pi+
@6
   name : Km
   smear : LHCbGeneric
   altMass : pi-
    \end{verbatim}
where the \verb!@! notation is used to number the particles in the decay file,
from left to right within each subdecay. Each particle can be given a unique {\tt name} that is used to
name the variables and histograms that are stored in the output data file. 
The effect of finite detector resolution is included through use of the {\tt smear}
keyword, typically only used for final state particles. By default, RapidSim includes
momentum smearing parameterisations that can be applied to the generated
four-vectors for pions, kaons,
protons, muons ({\tt LHCbGeneric}) or electrons ({\tt LHCbElectron}). It is simple for the user to
extend the available resolution parameterisations.
The {\tt altMass} keyword tells RapidSim to also compute the four-vector of the
considered particle under a particle mass hypothesis that is different to the default mass.
Appropriately named  variables are added to the output file.
The list of kinematic variables 
that RapidSim  calculates for each of the decays and for each final state particle are defined
at the head of the file using the \verb!paramsDecaying! and \verb!paramsStable! keywords.

\section{Heavy quark production}
\label{sec:Production}

\subsection{FONLL}
\label{sec:Production:fonll}

\begin{figure}[t]
\begin{center}
\includegraphics[width=0.49\textwidth]{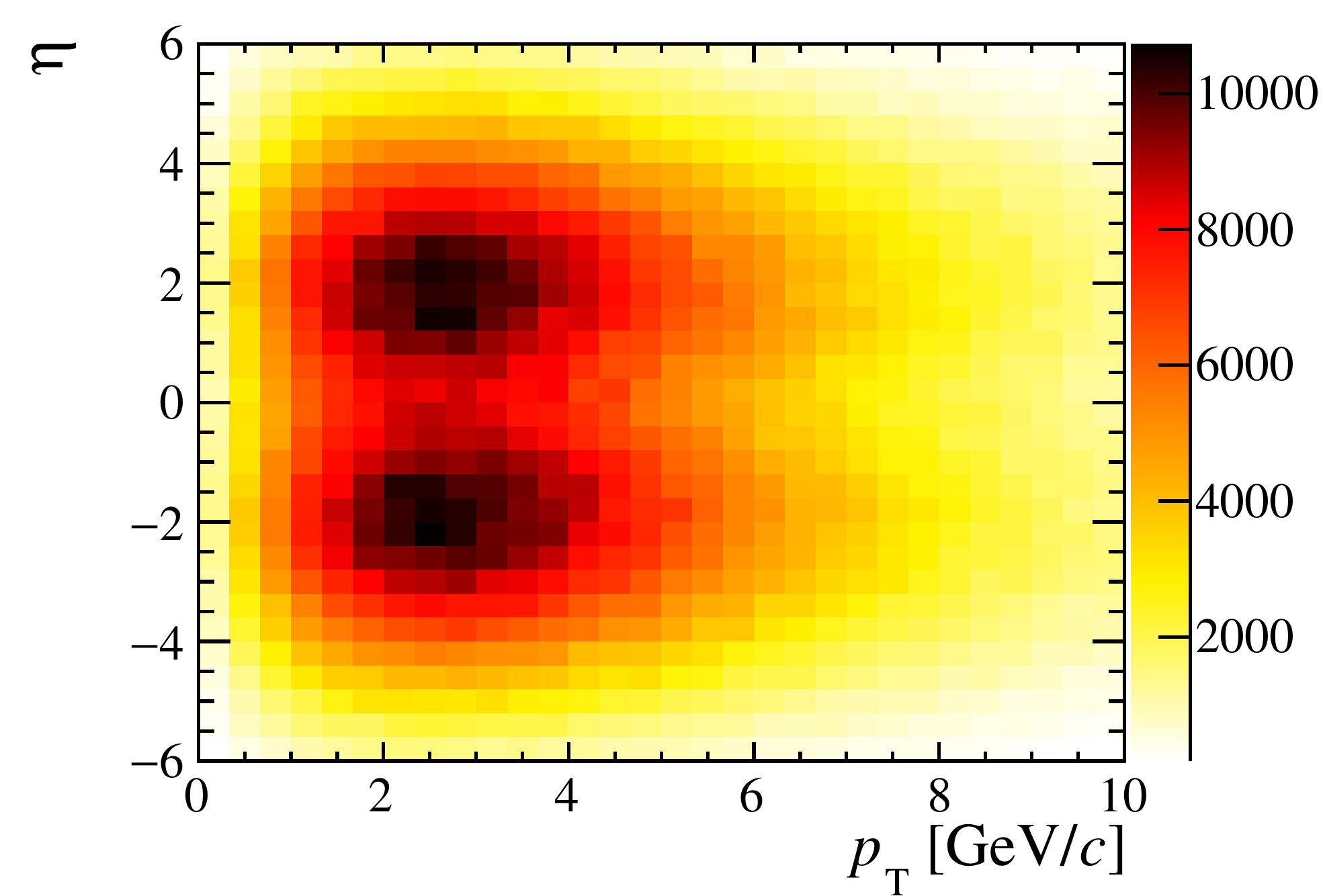}
\includegraphics[width=0.49\textwidth]{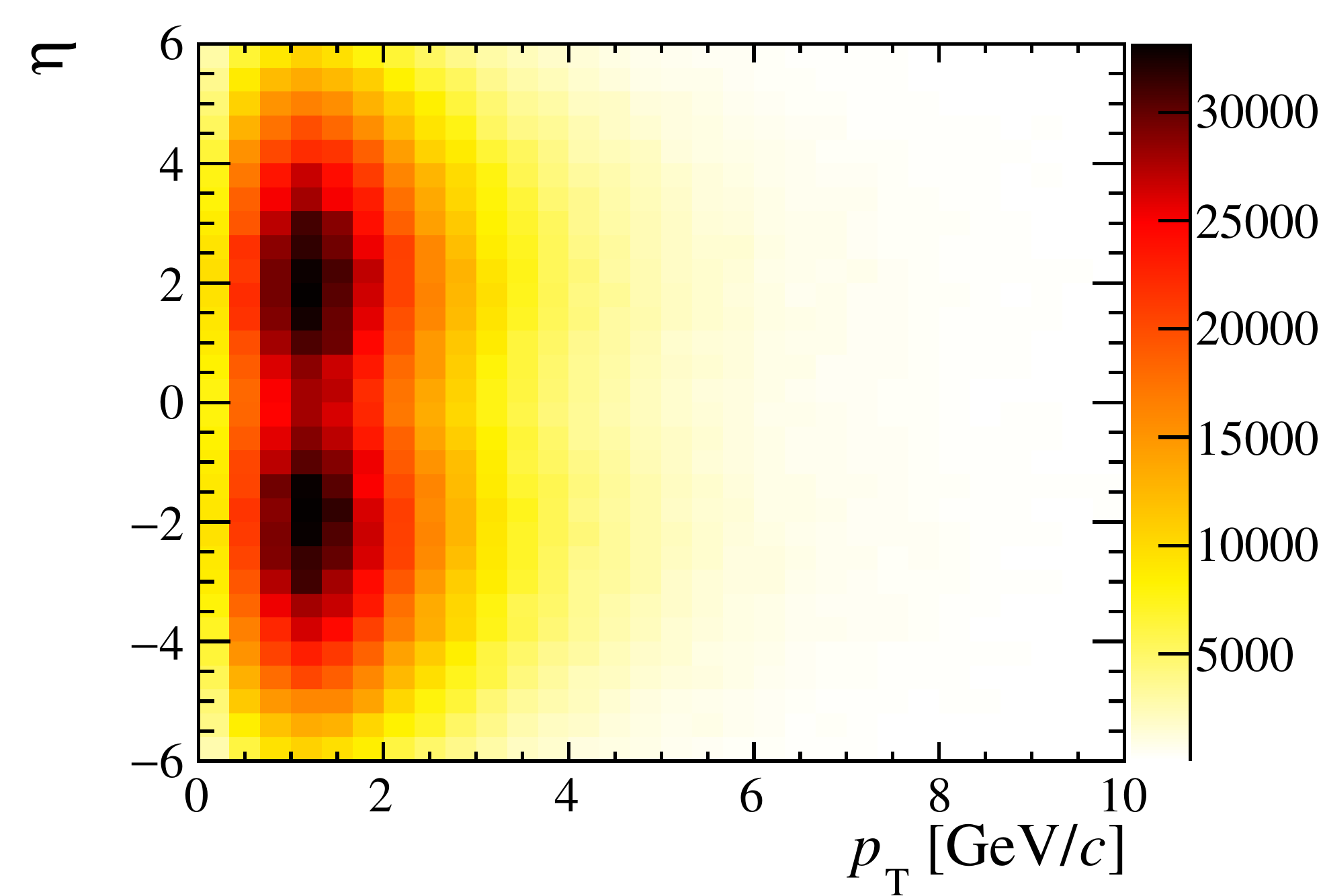}
\put(-300,120){(a)}
\put(-60,120){(b)}\\
\end{center}
\caption{\small 
Distribution in $\eta$ and $\pt$ of (a) $\Bs$ mesons and (b) $\Dz$ mesons produced in $7$ TeV $\proton\proton$ collisions.
}
\label{fig:prodfonll}
\end{figure}

In producing a realistic simulation of decays at a hadronic collider, it is important to determine the kinematic distribution of the particles produced. 
RapidSim generates particles following a particle-gun-like approach --- only the decaying particle is simulated, not the underlying event. 
This offers a significant time saving, but the kinematic distribution of the generated particles must be defined. 
RapidSim samples histograms generated from fixed-order next-to-leading-logarithm calculations~\cite{Cacciari:1998it} 
to determine the pseudorapidity, $\eta$, and transverse momentum, $\pt$, of the decaying particle. 
Example distributions are shown in Figure~\ref{fig:prodfonll}.

Histograms are included with the current version of RapidSim for the $4\pi$
production distributions of $\bquark$- and $\cquark$-hadrons in $\proton\proton$
collisions at a range of centre-of-mass energies. 
The distribution used is configured at run time using the global option {\tt energy} and the ID of the decaying particle. 
If necessary the {\tt mother} option may be used to override the choice of histogram. 

\subsubsection{Custom production kinematics}

Additional histograms may be saved to the {\tt rootfiles/fonll/} directory. Each file must contain two {\tt TH1} objects
named {\tt pT} and {\tt eta} describing the differential production cross sections in transverse momentum and pseudorapidity, respectively.
These may be obtained using the web interface provided at the URL in Ref~\cite{Cacciari:1998it}.
Files using the naming scheme {\tt LHC\{flavour\}\{energy\}.root} may be added without the need to recompile the executable. Extensions to the naming scheme must be made in the {\tt RapidConfig::loadParentKinematics} method.

\subsection{Non-phase space distributions}

\begin{figure}[t]
\begin{center}
\includegraphics[width=0.49\textwidth]{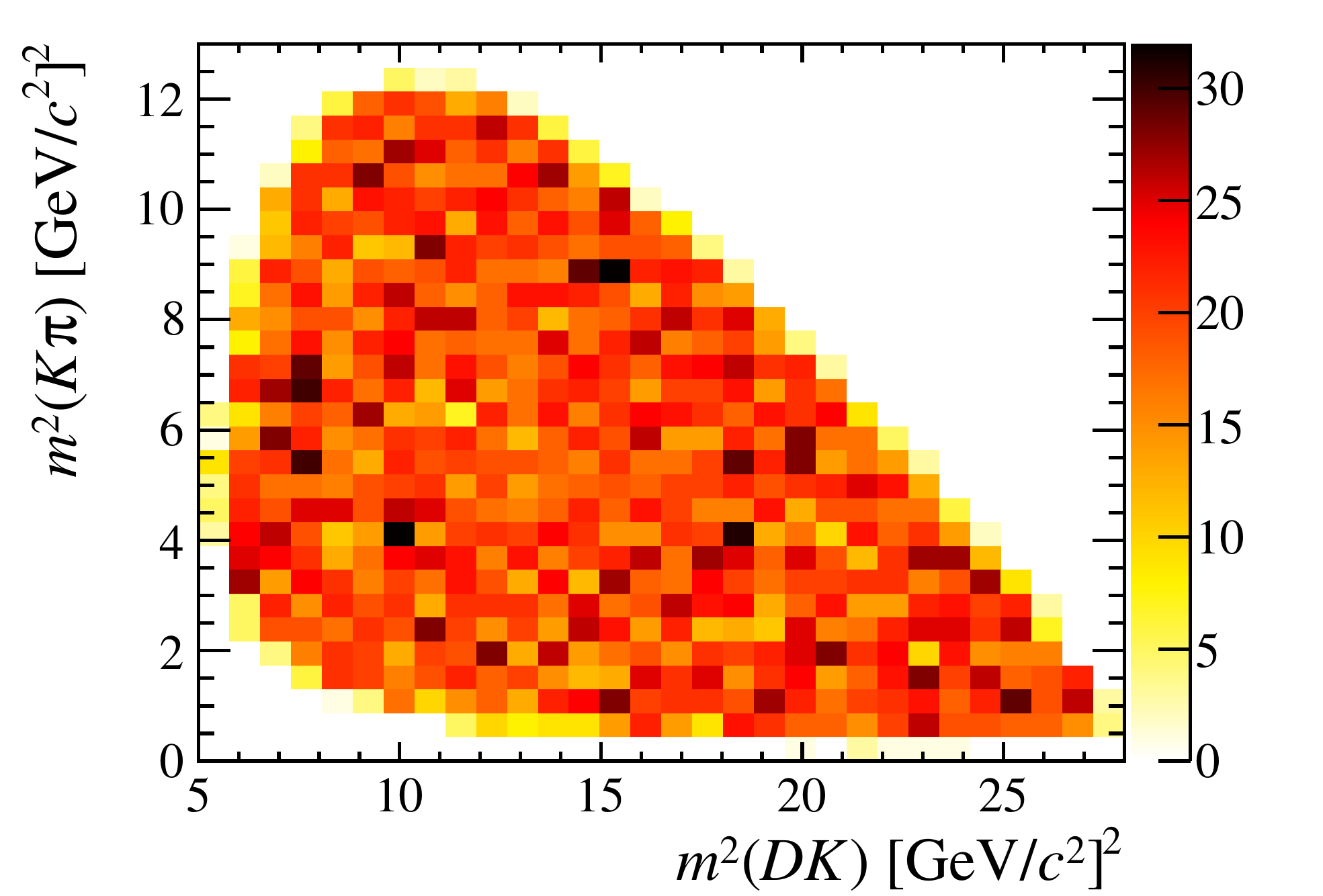}
\includegraphics[width=0.49\textwidth]{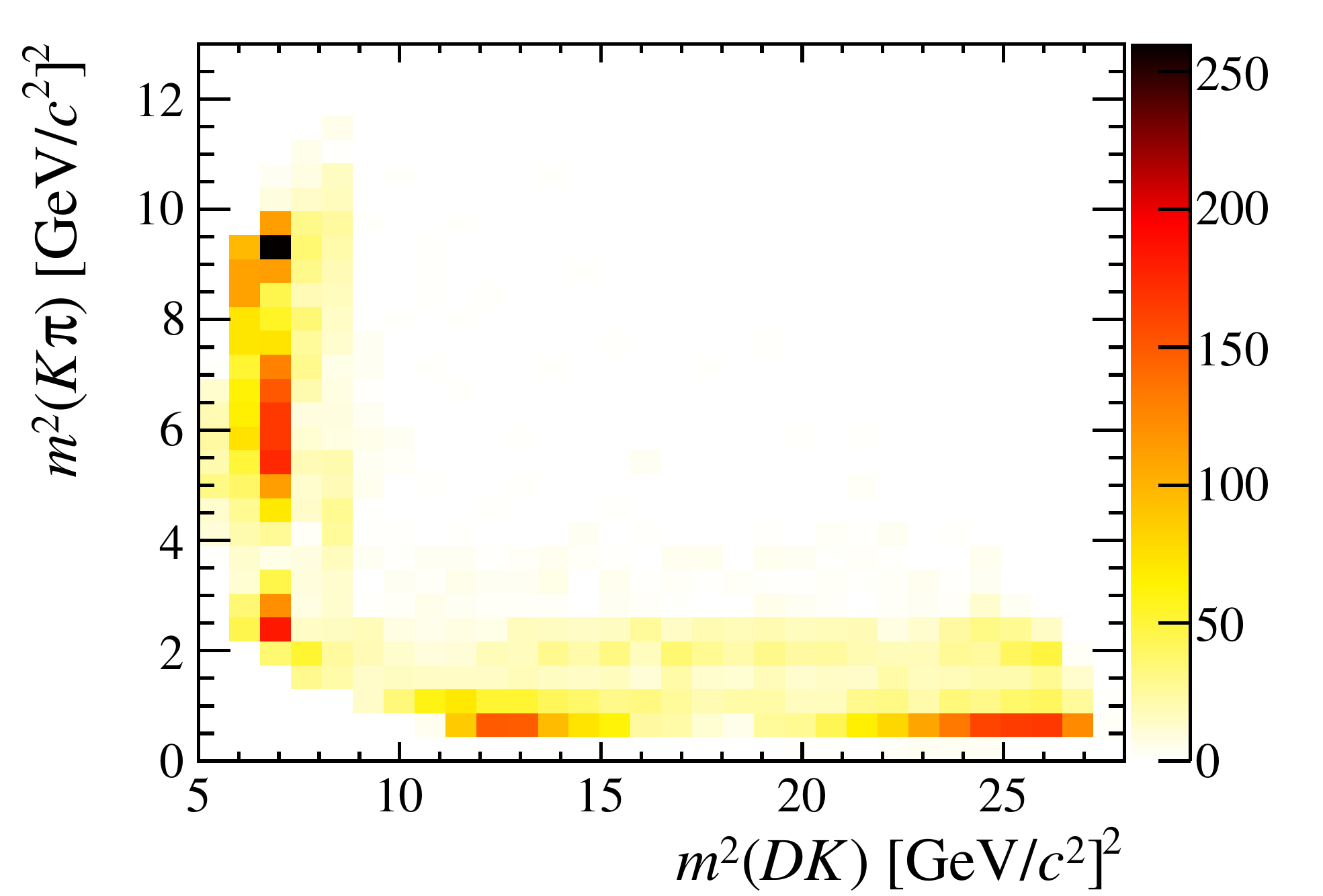}
\put(-300,120){(a)}
\put(-60,120){(b)}\\
\end{center}
\caption{\small 
Dalitz plots of $\Bs\to\Dzb\Km\pip$ (a) with a phase space distribution and (b) generated according to a 2D histogram.
}
\label{fig:prodshape}
\end{figure}

By default, RapidSim generates all decays flat across the available phase space,
however, in some cases it may be preferable to generate decays according to another distribution.
RapidSim features the option to generate decays according to a one or two dimensional
distribution in any defined parameters (see {\tt param} option in
Section~\ref{sec:Configuration}). This
is implemented via a Monte-Carlo accept/reject algorithm which is used to shape
the initially generated phase-space distribution.
Note that the requested number of events will be generated within the
specified distribution, so inefficient accept/reject functions (such as a Dalitz plot
with narrow resonances) will significantly increase the time required to generate each decay. 
An example is given in Figure~\ref{fig:prodshape}. 
This option is configured at runtime using the {\tt shape}
global option, \eg\ {\tt shape : myFile.root myHisto M2\_12 M2\_23}.

\section{Efficiency}
\label{sec:Efficiency}

RapidSim can account for two different types of efficiency: geometrical acceptance and selection requirements placed on kinematic quantities. 
These are detailed in Sections~\ref{sec:Efficiency:geom} and~\ref{sec:Efficiency:sel}, respectively.

\subsection{Geometrical acceptance}
\label{sec:Efficiency:geom}
\begin{figure}[t]
\begin{center}
\includegraphics[width=0.49\textwidth]{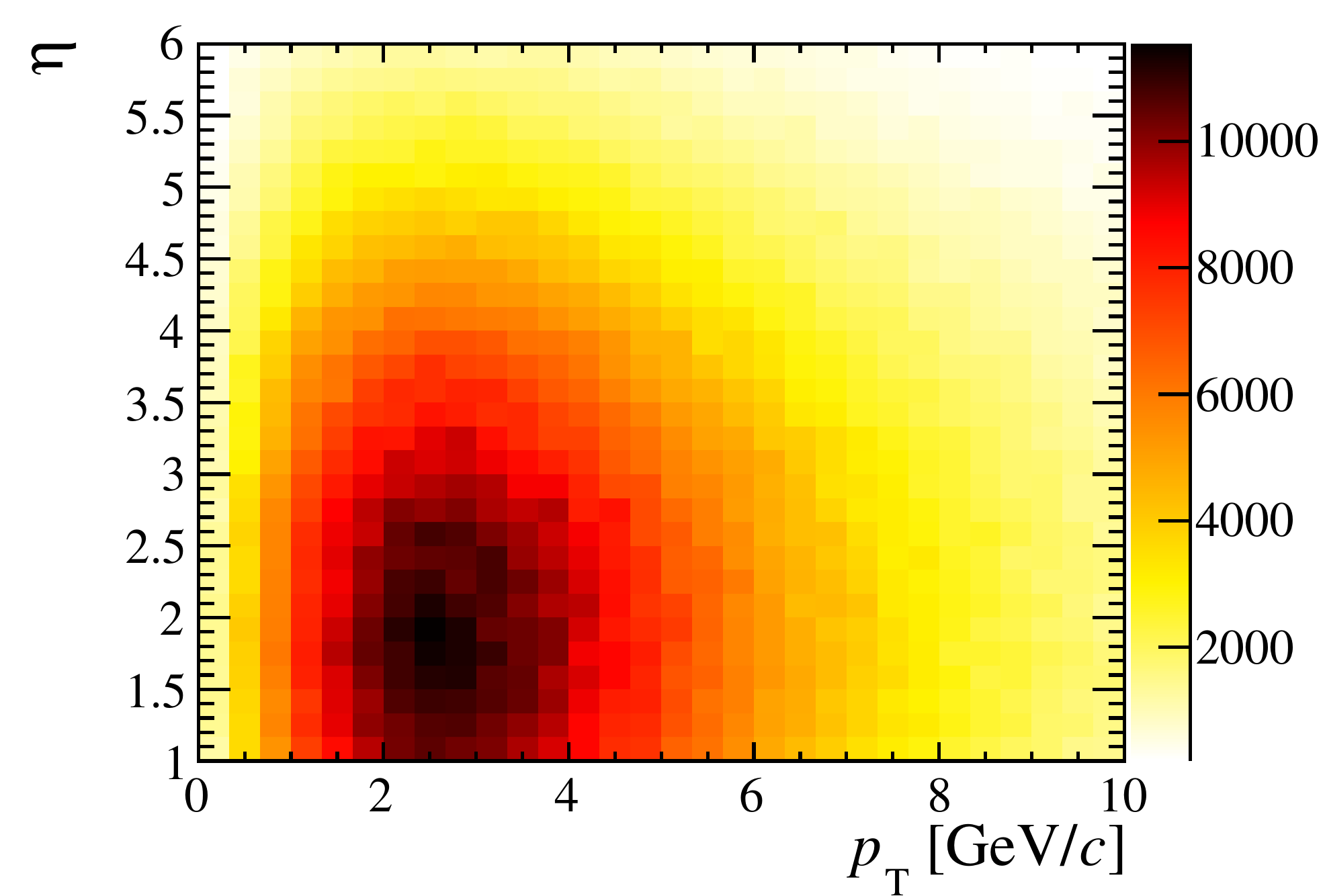}
\includegraphics[width=0.49\textwidth]{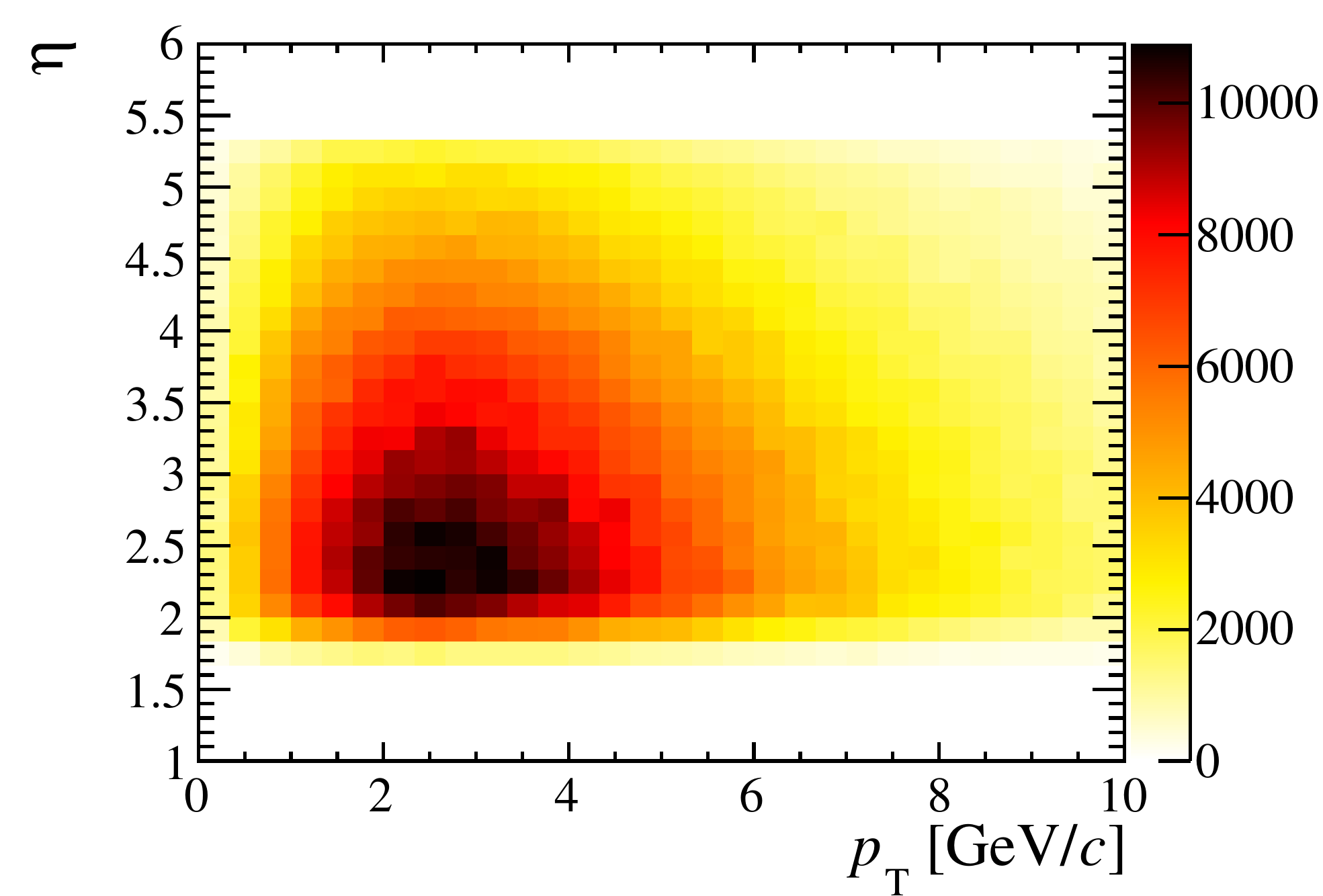}
\put(-300,120){(a)}
\put(-60,120){(b)}\\
\includegraphics[width=0.49\textwidth]{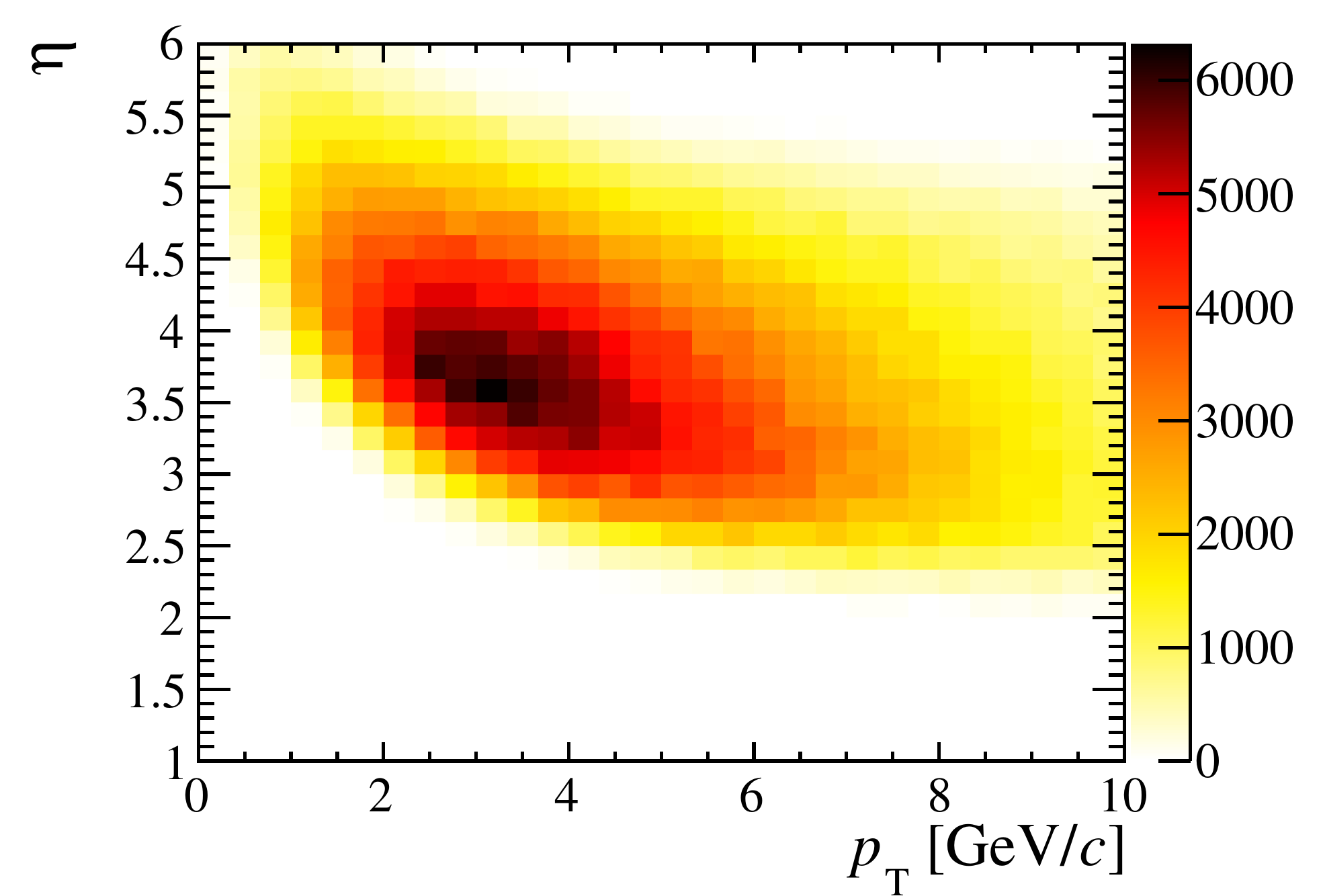}
\includegraphics[width=0.49\textwidth]{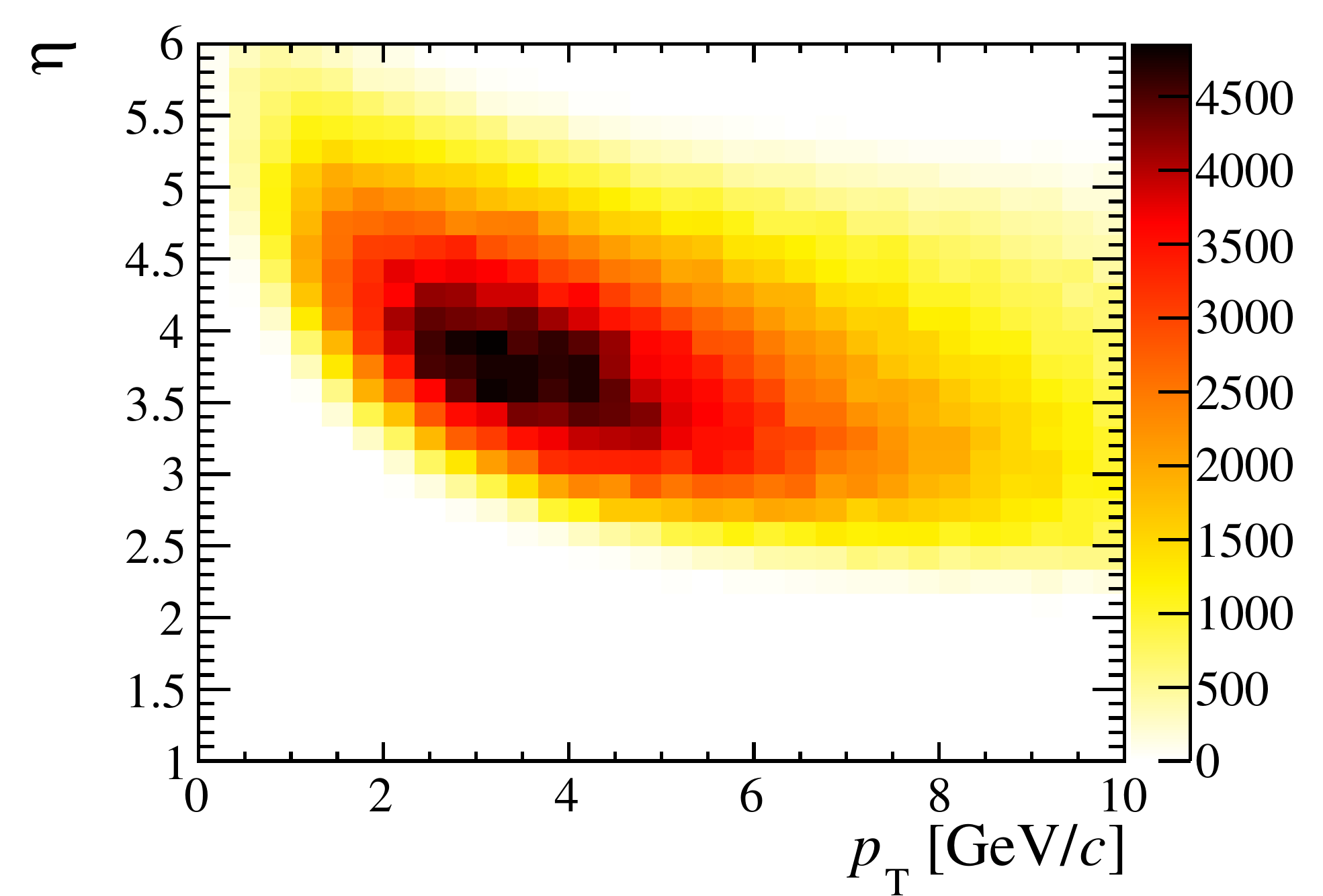}
\put(-300,120){(c)}
\put(-60,120){(d)}
\end{center}
\caption{\small 
Distribution in $\eta$ and $\pt$ of $\Bs$ mesons produced in $7$ TeV $\proton\proton$ collisions with (a) no acceptance requirement; 
(b) the $\Bs$ meson required to be within the LHCb acceptance; (c) the decay products (from $\Bs\to\jpsi(\to\mumu)\phi(\to\Kp\Km)$ decay) 
required to be within the LHCb acceptance; and (d) the decay products required to be within the LHCb acceptance downstream of the dipole magnet. 
\nb\ by default, RapidSim only generates decays in the forward region ($1.0 < \eta < 6.0$) when using the LHCb geometry.
}
\label{fig:effgeom}
\end{figure}

A requirement may be placed on the geometry of the decay. 
This may be configured via the configuration file at run time from a selection of defined options,
however, new options must first be defined at compile time within the {\tt RapidAcceptance} class.
In the current version of the package, three LHCb-specific options are available in addition to the default:
\begin{description}
	\item [Any] No requirement is applied;
	\item [MotherIn] The decaying particle is required to be within the angular acceptance of the LHCb detector;
	\item [AllIn] All of the final-state particles are required to be within the angular acceptance of the LHCb detector;
	\item [AllDownstream] All of the final-state particles are required to remain in the acceptance of the LHCb detector downstream of the dipole magnet.
\end{description}
The single acceptance requirement is configured using the global option {\tt acceptance}, \eg\ {\tt acceptance : Any}.
Figure~\ref{fig:effgeom} shows an example of the distribution of transverse momentum and pseudorapidity for \Bs
mesons with each of the acceptance requirements applied. 

\subsection{Kinematic requirements}
\label{sec:Efficiency:sel}
\begin{figure}[t]
\begin{center}
\includegraphics[width=0.49\textwidth]{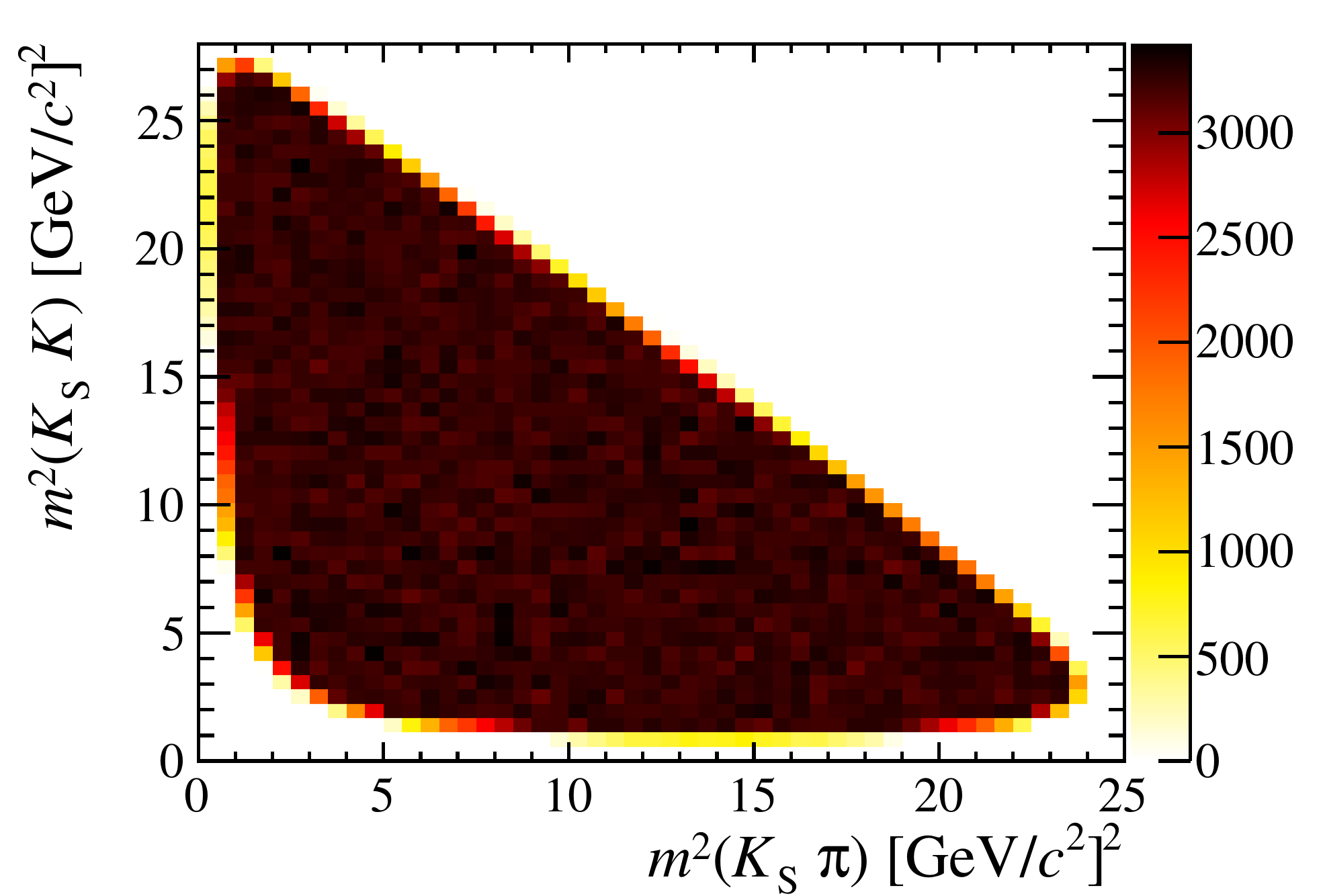}
\includegraphics[width=0.49\textwidth]{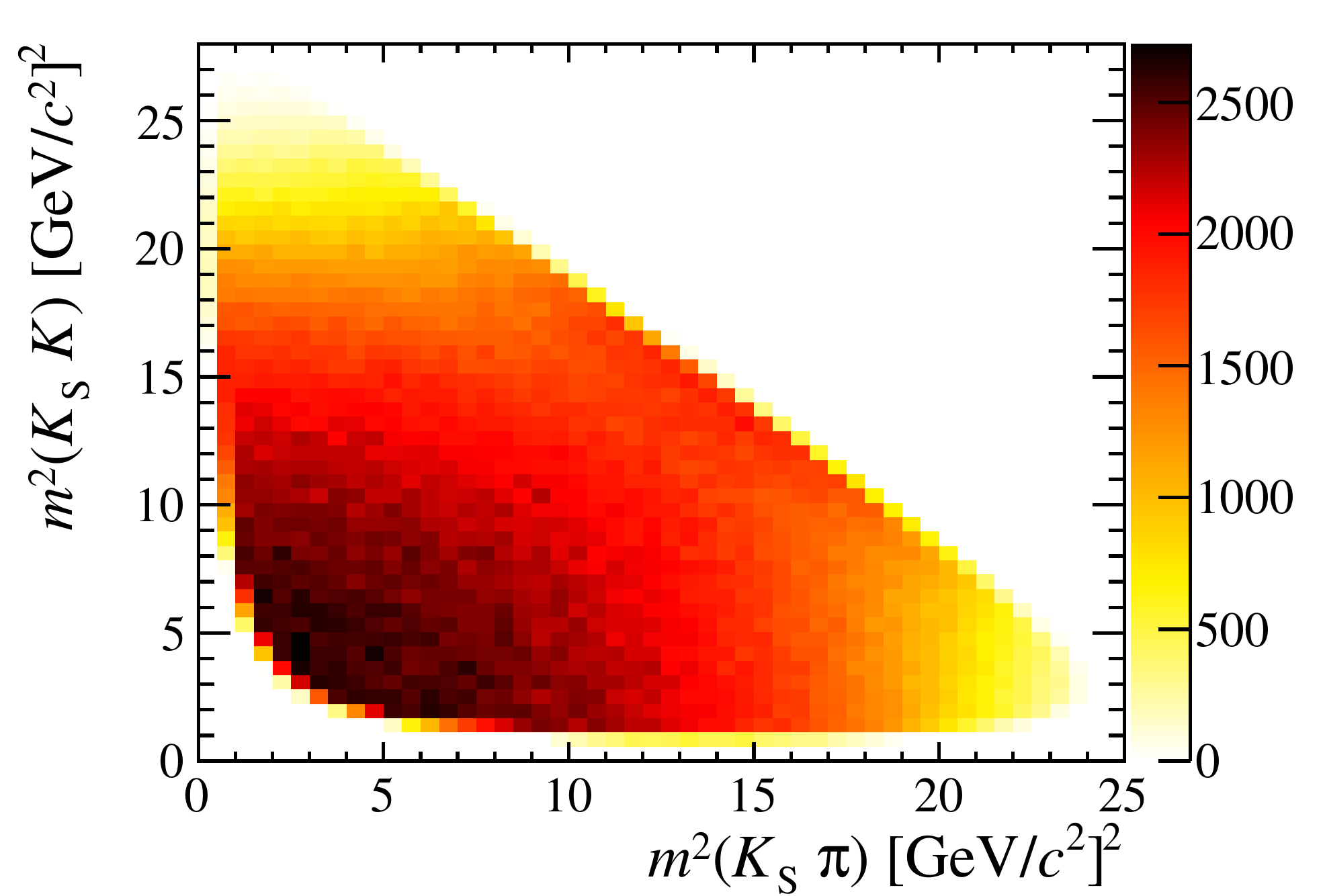}
\put(-300,120){(a)}
\put(-60,120){(b)}
\end{center}
\caption{\small Example phase space Dalitz plot ($\Bs\to\KS \Kp \pi^-$) with (a) no selection requirements and (b) $\pt$ requirements on the kaon and pion to be larger than 1 \gevcc.}
\label{fig:effsel}
\end{figure}

Selection requirements may be placed on any defined parameter (see {\tt param} option
in Section~\ref{sec:Configuration}). 
Requirements may be a minimum ({\tt min}); a maximum ({\tt max}); a range ({\tt range}); or an excluded range ({\tt veto}) and are configured using the global option {\tt cut}, \eg\ {\tt cut : pion\_PT range 1 100}. 
An example is shown in Figure~\ref{fig:effsel}.

\subsection{Missing particles}
\label{sec:efficiency:missing}
\begin{figure}[t]
\begin{center}
\includegraphics[width=0.49\textwidth]{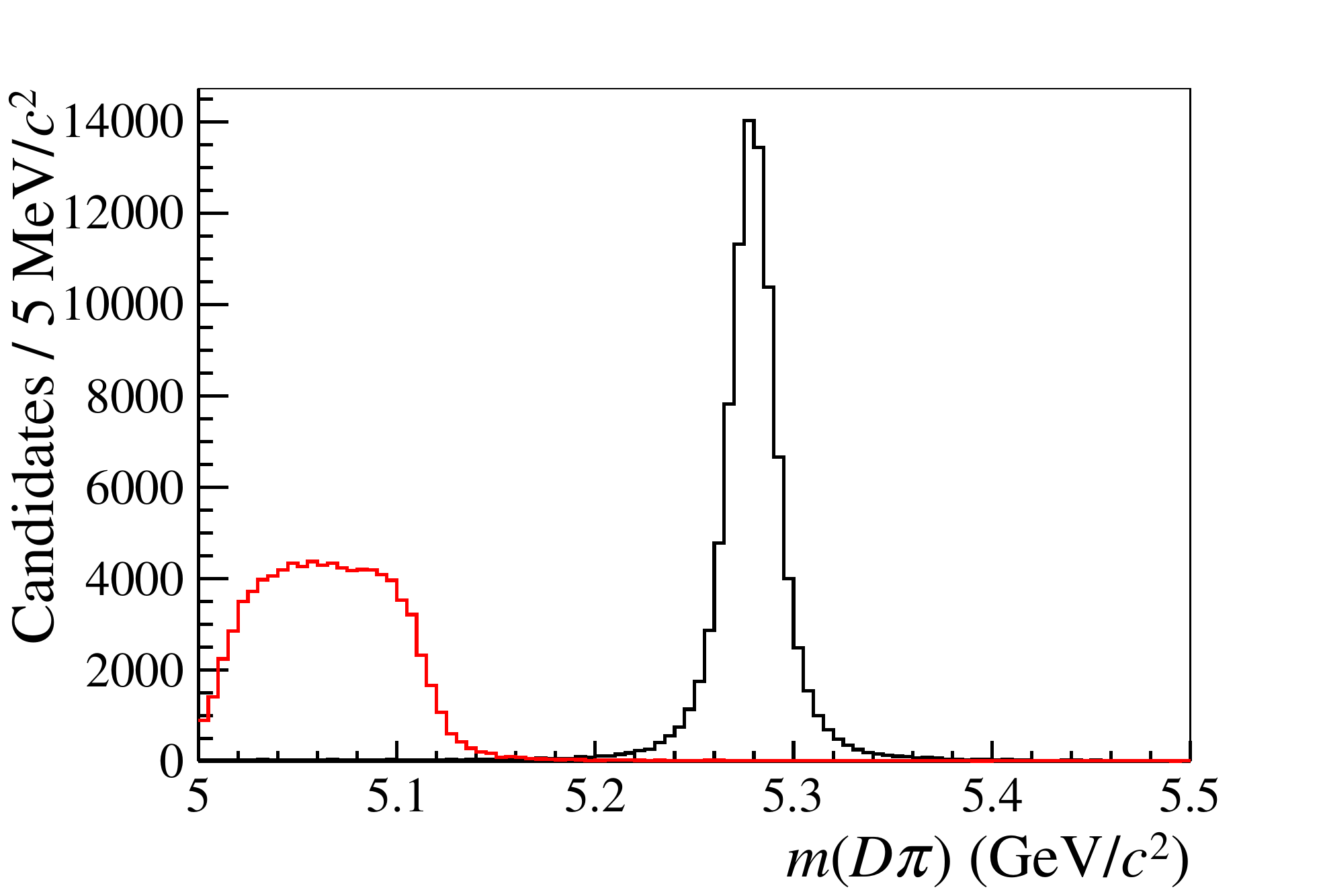}
\end{center}
\caption{\small Distribution of $m(D\pi)$ invariant mass for (black) $\Dz\pip$ decays where all particles are
reconstructed and (red) $\Dstarz\pip$ decays where a neutral pion is missed.}
\label{fig:missing}
\end{figure}

In certain decays some of the stable particles are not reconstructed
(\eg\ neutrinos in semi-leptonic decays or missing particles in partially reconstructed backgrounds). 
In such cases, it is of interest to reconstruct the mass or momentum of the parent particle excluding
the momentum lost to the missing child particle. 
This may be reproduced in RapidSim using the {\tt invisible} option.
This option is turned on by default for neutrinos but may also be used for any other particle type. 
This option only affects reconstructed quantities --- true quantities will account for the invisible particle. 
An example of this option is shown in Figure~\ref{fig:missing}.

\clearpage
\section{Momentum smearing}
\label{sec:Smearing}

\begin{figure}[t]
\begin{center}
\includegraphics[width=0.49\textwidth]{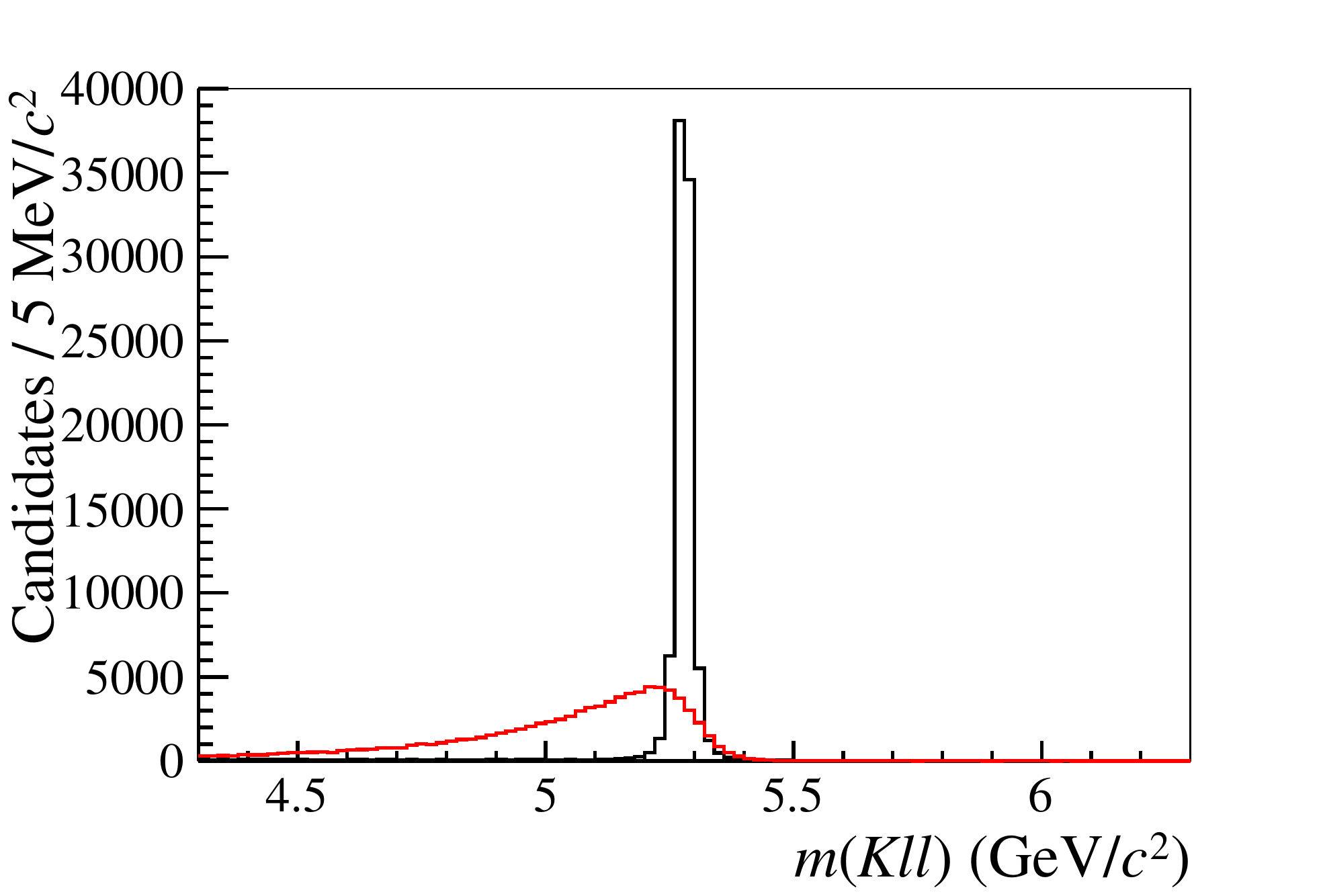}
\end{center}
\caption{\small Simulated resolution effects on the distribution of $m(\Kp l^+l^-)$ for \Bp mesons decaying to (black) $\Kp\mumu$ and (red) $\Kp\ep\en$. The {\tt LHCbGeneric} smearing option has been used for the kaon and muons, while the {\tt LHCbElectron} option (which accounts for Bremsstrahlung) has been used for the electron and positron.}
\label{fig:smearing}
\end{figure}

RapidSim simulates the effect of a finite momentum resolution by smearing the momentum of each stable particle according to a defined resolution function. 
Smearing options are defined in the {\tt config/smear/} directory (as described in Section~\ref{sec:Smearing:config})
and configured for each particle using the {\tt smear} keyword. 
Resolution functions may be defined either by a Gaussian distribution
with a $\ptot$-dependent (or $\pt$- and $\eta$-dependent) width, or by 
a series of histograms that describe the resolution in different $\ptot$ ranges. 
The former approach allows the width of the resolution function
to vary continuously and is the best option in most cases. 
The predefined {\tt LHCbGeneric} smearing option uses this approach to model the resolution effects on 
charged hadrons and muons in the LHCb experiment, based on the calibration documented in
Refs.~\cite{Needham:2007zz,Aaij:2013uaa}. 
The histogram-based approach allows for asymmetric effects such as Bremsstrahlung to be modelled and 
is used in the predefined {\tt LHCbElectron} option, which is based on a modified version of the model
given in the supplementary material of Ref.~\cite{Aaij:2013cby}.
Examples of these two smearing options are shown in Figure~\ref{fig:smearing}.
The {\tt AtlasHadron} and {\tt AtlasMuon} options use the momentum resolution calibration
from Ref.~\cite{ATLAS-CONF-2010-009} to model the resolution effects as a function
of $\pt$ and $\eta$ in the ATLAS experiment for hadrons and muons.
It is assumed that the resolution for hadrons is uniform as a function of \pt, which is reasonable
for tracks with $\pt>1 \gevc$.

\begin{figure}[t]
\begin{center}
\includegraphics[width=0.49\textwidth]{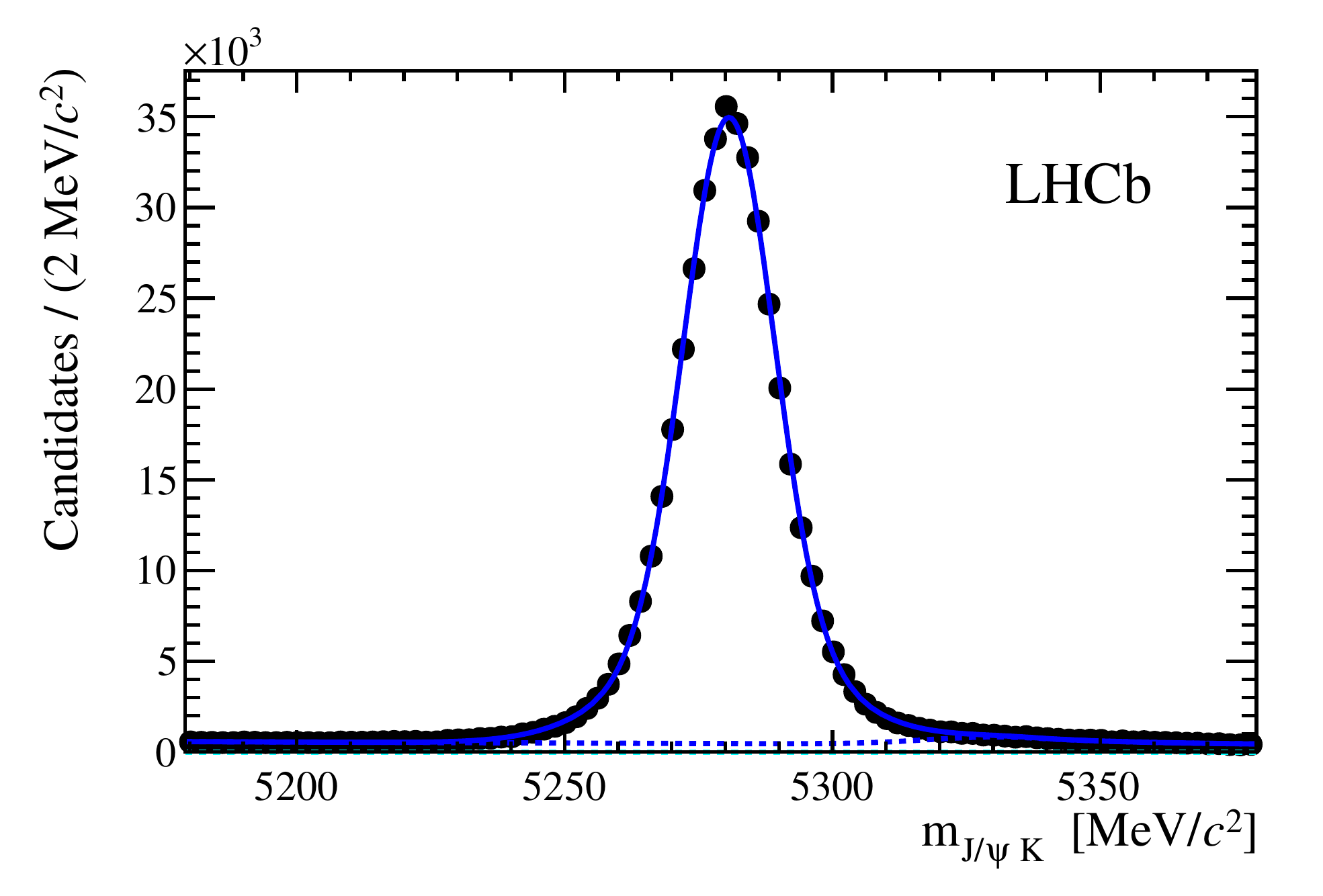}
\includegraphics[width=0.49\textwidth]{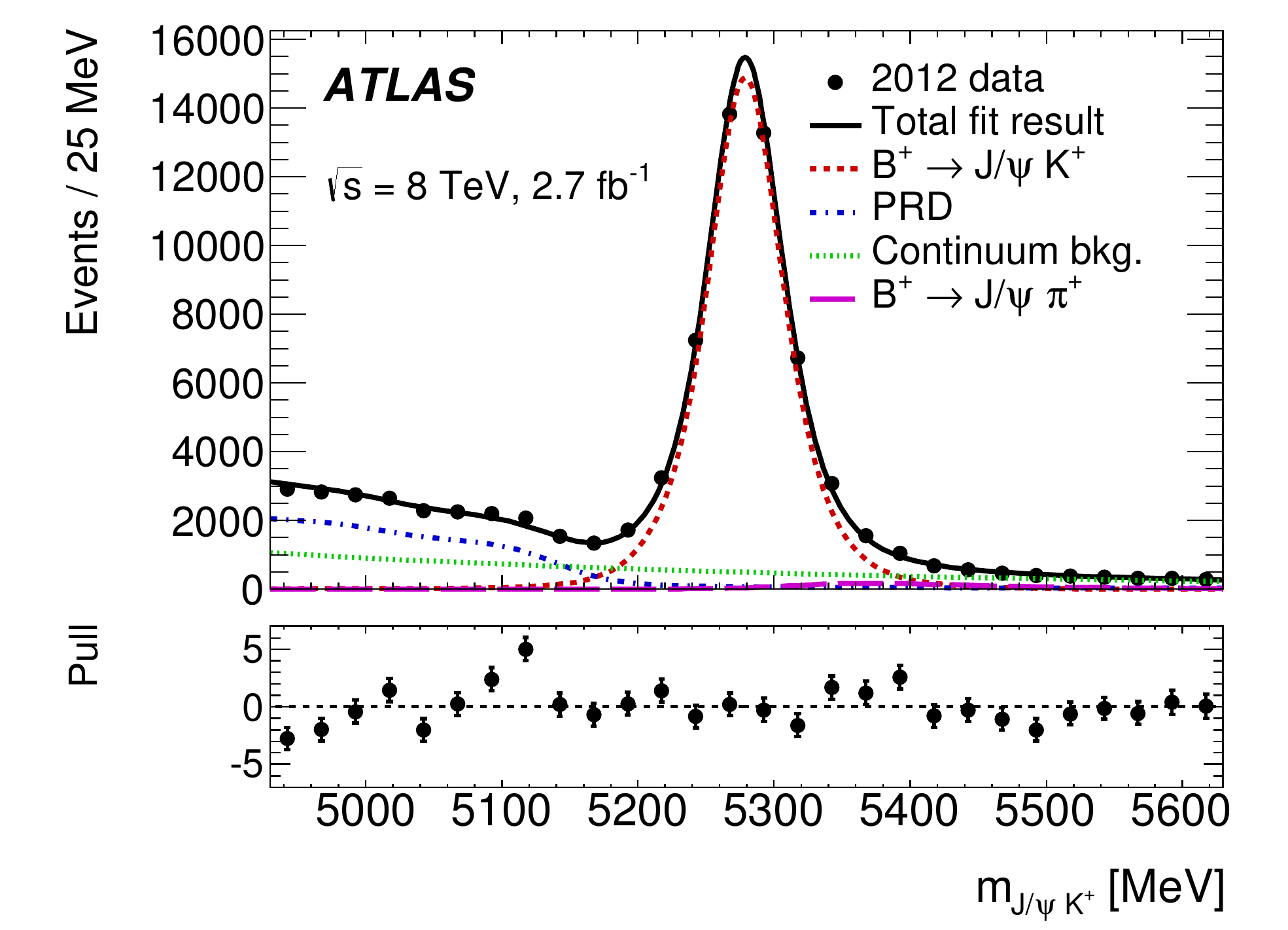}
\put(-300,100){(a)}
\put(-60,80){(b)}\\
\includegraphics[width=0.49\textwidth]{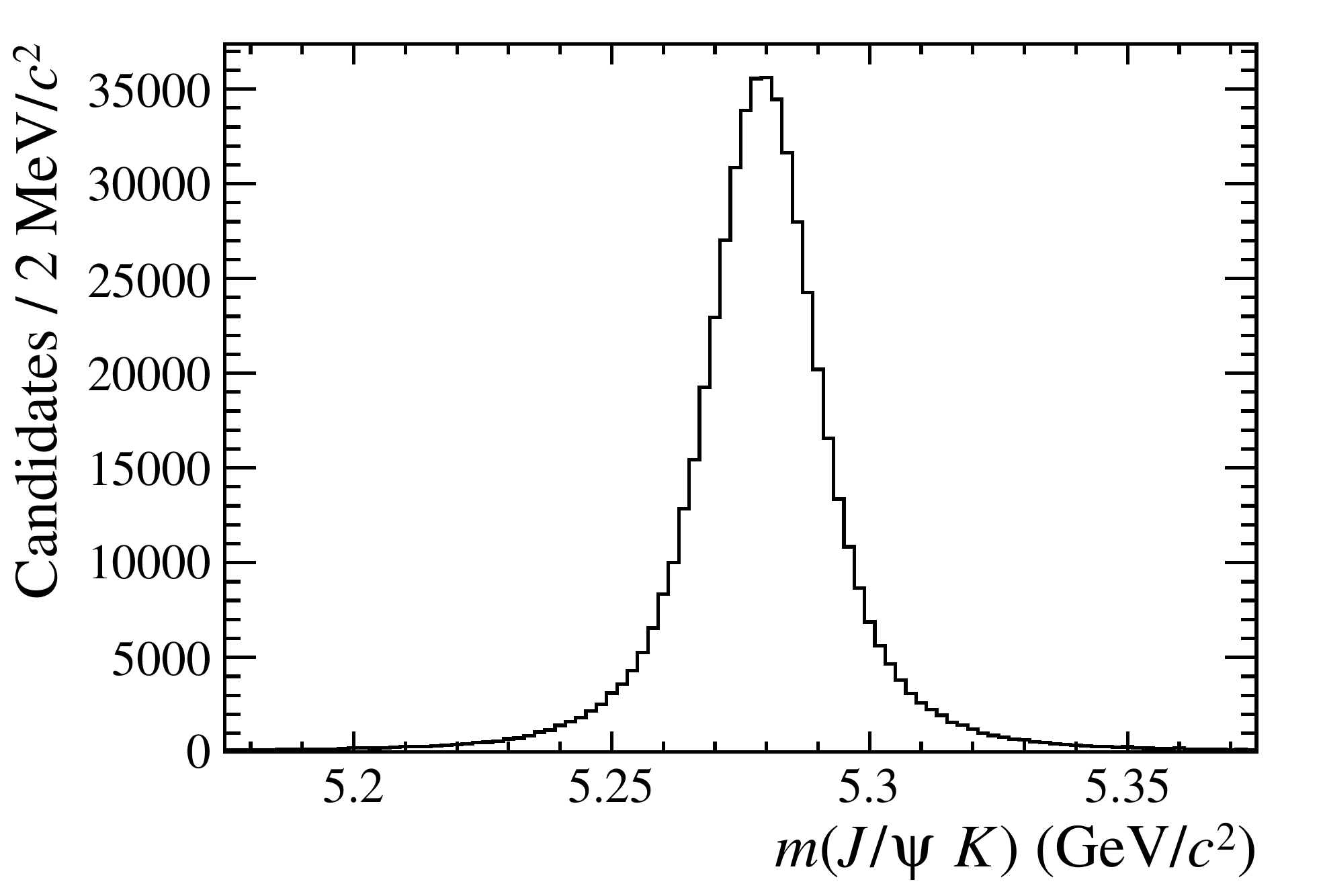}
\includegraphics[width=0.49\textwidth]{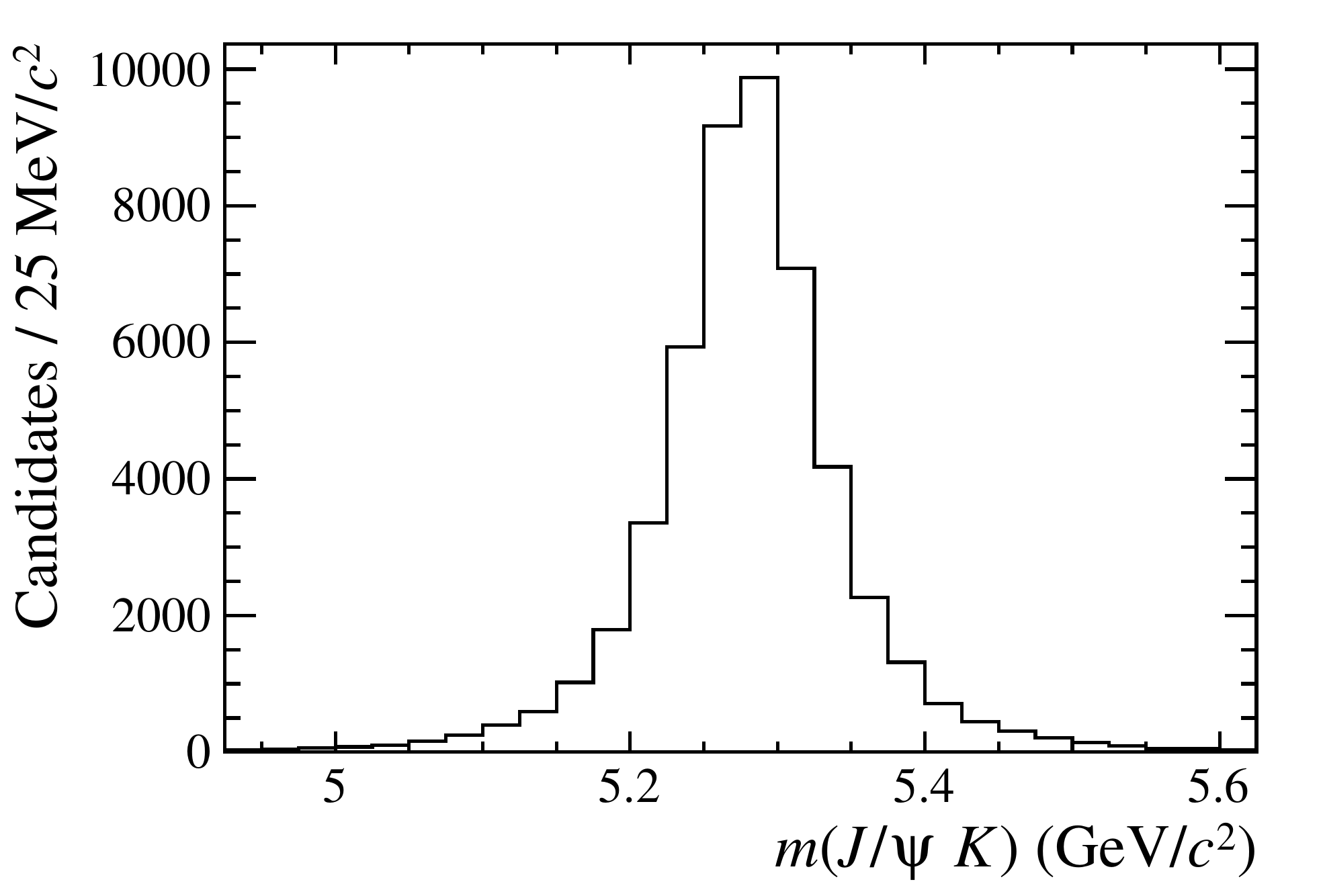}
\put(-300,100){(c)}
\put(-60,100){(d)}
\end{center}
\caption{\small Distributions of the invariant mass of $\Bu\to\jpsi\Kp$ decays in the (a) LHCb~\cite{Aaij:2012nna}
and (b) ATLAS~\cite{Aaboud:2016ire} experiments and those generated by RapidSim for (c) LHCb conditions
and (d) ATLAS conditions. Note the different scales on the abscissas.}
\label{fig:Bu2JpsiK}
\end{figure}

Figure~\ref{fig:Bu2JpsiK} shows examples of the distributions of 
$\jpsi\Kp$ invariant mass in selected $\Bu\to\jpsi\Kp$ decays
from the LHCb and ATLAS experiments and the corresponding
signal-only distributions generated using RapidSim with the relevant momentum smearing options
enabled.
The RapidSim simulation manages to reproduce the mass resolution observed in data.

\subsection{Configuration}
\label{sec:Smearing:config}

Each smearing option is defined by a single ASCII configuration file and a single ROOT file 
containing either a {\tt TGraph} object (for Gaussian resolution options with a $\ptot$-dependent width),
a {\tt TH2} object (for Gaussian resolutions with $\pt$ and $\eta$ dependence), 
or one or more {\tt TH1} objects (for histogram-based options). 
The configuration file is formatted as follows:
\begin{itemize}
    \item The first line gives the path to the associated ROOT file.
    \item The second line is {\tt GAUSS} for $\ptot$-dependent Gaussian smearing options, {\tt GAUSSPTETA} for $\pt$- and $\eta$-dependent Gaussian smearing options, 
          and {\tt HISTS} for histogram smearing options.
    \item For Gaussian smearing options, the third line gives the name of the {\tt TGraph} ({\tt TH2}) object that gives the resolution width as a function of $\ptot$ ($\pt$ and $\eta$).
    \item For histogram smearing options, the third and subsequent lines give the lower edge of a $\ptot$ bin and the name of the {\tt TH1} object describing the resolution function in that bin. 
	  An example file is given below.
\end{itemize}
\begin{verbatim}
electronHists.root
HISTS
0       P0_hist
10000   P1_hist
15000   P2_hist
20000   P3_hist
30000   P4_hist
50000   P5_hist
\end{verbatim}

\section{Particle mis-identification}
\label{sec:Misid}

\begin{figure}[t]
\begin{center}
\includegraphics[width=0.49\linewidth]{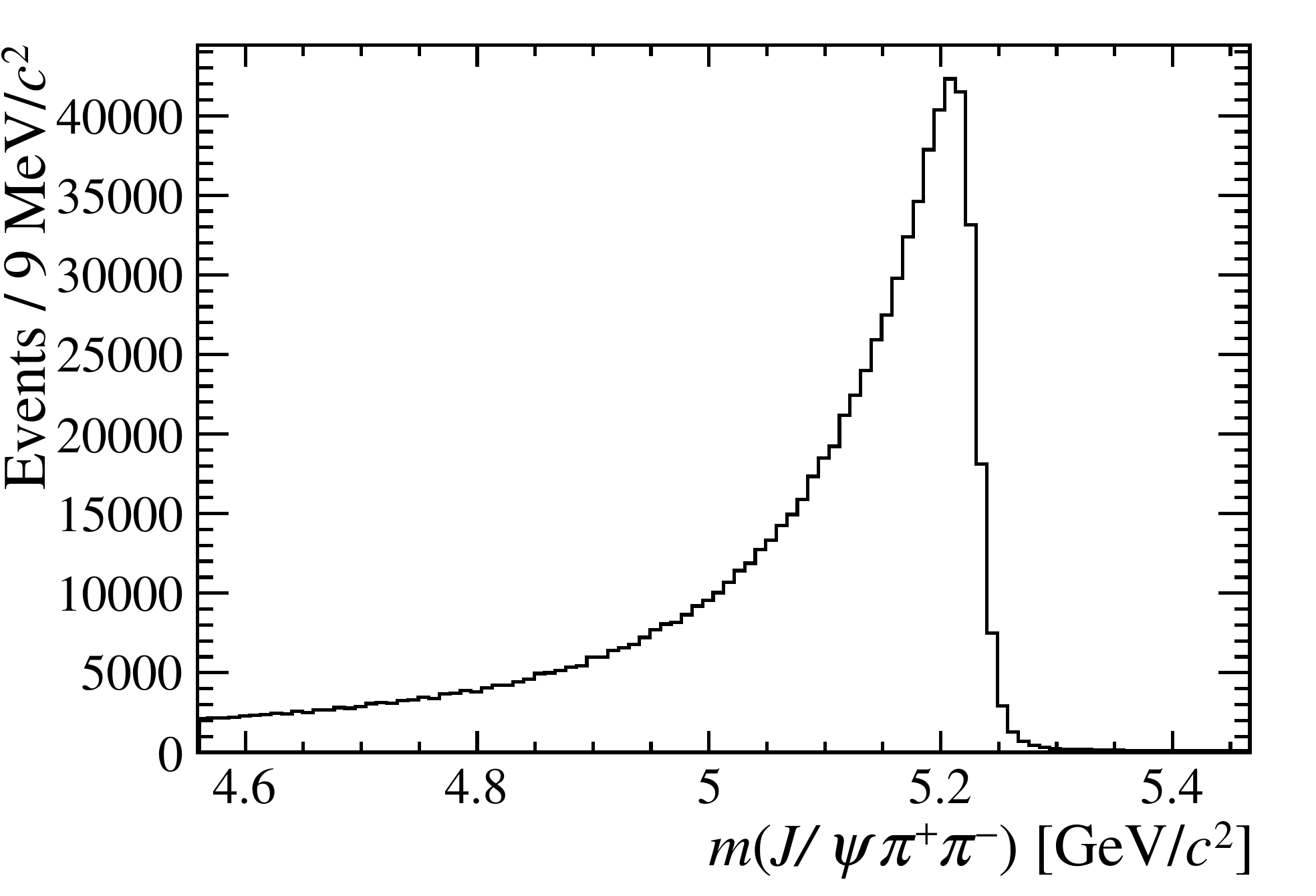}
\end{center}
\caption{\small Distribution of $\jpsi\pi^+\pi^-$ invariant mass for $\Bs\to\jpsi\phi$ decays where both kaons have been mis-identified as pions.}
\label{fig:misid}
\end{figure}

RapidSim can also calculate quantities using an alternative mass hypothesis for one or two particles. 
This is particularly useful when investigating backgrounds where a particle has been mis-identified. 
Multiple alternative mass hypotheses may be defined for each particle using the particle's {\tt altMass}
option (\eg\ {\tt altMass : K+ mu+} will add both kaon and muon masses as alternative hypotheses). 
All parameters will be calculated for all combinations of hypotheses with two or fewer mis-identified particles. 
For instance, $\Dz\to\Km\pip$ with $\pi\to\mu$, $\pi\to K$ and $K\to\pi$ mis-identifications considered will result in a total of five alternative mass hypotheses ($\Km\mup$, $\Kp\Km$, $\pim\pip$, $\pim\mup$ and $\pim\Kp$). 
Figure~\ref{fig:misid} shows an example mass distribution calculated with
mis-identifications of two final state particles.

\section{Summary}
\label{sec:Conclusion}

A common task in the analysis of heavy-flavour hadron decays at particle
physics experiments is to understand the kinematic properties of the signal and background
processes, in particular when determining selection requirements or when describing the
invariant mass spectrum of the final state particles. RapidSim is a new lightweight
application that allows the analyst to quickly generate specific heavy-flavour hadron
decays with realistic production kinematic distributions, efficiencies and momentum
resolutions. The specific decay properties and options are controlled via simple user-defined
configuration files. The production environment,  momentum resolutions
and desired efficiency are defined via externally provided parameterisations. The
simple framework allows the user to easily extend the simulation should more detail
be required, such as the addition of alternative detector geometries
beyond the defaults provided. The problem of understanding the shape of hadron decays
where one or more of the final state particles are mis-identified is handled in a simple
way via a configuration file. Altogether, RapidSim packages many aspects of 
typical particle physics analyses into a simple application that gives users the 
ability to rapidly generate large samples of decays that can be used to aid their
understanding without the need to generate fully simulated decays
and the corresponding use of computing and storage resources that it entails.
It is already being utilised in ongoing analyses within the LHCb collaboration.

\section*{Acknowledgements}

We would like to thank the support of the Science and Technology Facilities Council grants
ST/K004646/1 and ST/M004058/1.

\addcontentsline{toc}{section}{References}
\setboolean{inbibliography}{true}
\bibliographystyle{LHCb}
\bibliography{main,LHCb-PAPER,LHCb-CONF,LHCb-DP,LHCb-TDR}

\ifx\mcitethebibliography\mciteundefinedmacro
\PackageError{LHCb.bst}{mciteplus.sty has not been loaded}
{This bibstyle requires the use of the mciteplus package.}\fi
\providecommand{\href}[2]{#2}
\begin{mcitethebibliography}{10}
\mciteSetBstSublistMode{n}
\mciteSetBstMaxWidthForm{subitem}{\alph{mcitesubitemcount})}
\mciteSetBstSublistLabelBeginEnd{\mcitemaxwidthsubitemform\space}
{\relax}{\relax}

\bibitem{g_a_cowan_2016_160402}
G.~A. Cowan, D.~Craik, and M.~Needham,
  \ifthenelse{\boolean{articletitles}}{\emph{gcowan/rapidsim: Rapidsim
  pre-release}, }{} Oct., 2016.
\newblock
  doi:~\href{http://dx.doi.org/10.5281/zenodo.160402}{10.5281/zenodo.160402}\relax
\mciteBstWouldAddEndPuncttrue
\mciteSetBstMidEndSepPunct{\mcitedefaultmidpunct}
{\mcitedefaultendpunct}{\mcitedefaultseppunct}\relax
\EndOfBibitem
\bibitem{Brun:1997pa}
R.~Brun and F.~Rademakers, \ifthenelse{\boolean{articletitles}}{\emph{{ROOT: An
  object oriented data analysis framework}},
  }{}\href{http://dx.doi.org/10.1016/S0168-9002(97)00048-X}{Nucl.\ Instrum.\
  Meth.\  \textbf{A389} (1997) 81}\relax
\mciteBstWouldAddEndPuncttrue
\mciteSetBstMidEndSepPunct{\mcitedefaultmidpunct}
{\mcitedefaultendpunct}{\mcitedefaultseppunct}\relax
\EndOfBibitem
\bibitem{Cacciari:1998it}
M.~Cacciari, M.~Greco, and P.~Nason,
  \ifthenelse{\boolean{articletitles}}{\emph{{The P(T) spectrum in heavy flavor
  hadroproduction}},
  }{}\href{http://dx.doi.org/10.1088/1126-6708/1998/05/007}{JHEP \textbf{05}
  (1998) 007}, \href{http://arxiv.org/abs/hep-ph/9803400}{{\normalfont\ttfamily
  arXiv:hep-ph/9803400}}\relax
\mciteBstWouldAddEndPuncttrue
\mciteSetBstMidEndSepPunct{\mcitedefaultmidpunct}
{\mcitedefaultendpunct}{\mcitedefaultseppunct}\relax
\EndOfBibitem
\bibitem{Evans:2008zzb}
L.~Evans and P.~Bryant, \ifthenelse{\boolean{articletitles}}{\emph{{LHC
  Machine}}, }{}\href{http://dx.doi.org/10.1088/1748-0221/3/08/S08001}{JINST
  \textbf{3} (2008) S08001}\relax
\mciteBstWouldAddEndPuncttrue
\mciteSetBstMidEndSepPunct{\mcitedefaultmidpunct}
{\mcitedefaultendpunct}{\mcitedefaultseppunct}\relax
\EndOfBibitem
\bibitem{Alves:2008zz}
LHCb collaboration, A.~A. Alves, Jr.\ {\em et~al.},
  \ifthenelse{\boolean{articletitles}}{\emph{{The LHCb Detector at the LHC}},
  }{}\href{http://dx.doi.org/10.1088/1748-0221/3/08/S08005}{JINST \textbf{3}
  (2008) S08005}\relax
\mciteBstWouldAddEndPuncttrue
\mciteSetBstMidEndSepPunct{\mcitedefaultmidpunct}
{\mcitedefaultendpunct}{\mcitedefaultseppunct}\relax
\EndOfBibitem
\bibitem{Needham:2007zz}
M.~Needham, \ifthenelse{\boolean{articletitles}}{\emph{{Performance of the LHCb
  track reconstruction software}}, }{} Tech. Rep. CERN-LHCB-2007-144,
  LPHE-2008-01, EPFL, 2007\relax
\mciteBstWouldAddEndPuncttrue
\mciteSetBstMidEndSepPunct{\mcitedefaultmidpunct}
{\mcitedefaultendpunct}{\mcitedefaultseppunct}\relax
\EndOfBibitem
\bibitem{Aaij:2013uaa}
LHCb collaboration, R.~Aaij {\em et~al.},
  \ifthenelse{\boolean{articletitles}}{\emph{{Precision measurement of D meson
  mass differences}}, }{}\href{http://dx.doi.org/10.1007/JHEP06(2013)065}{JHEP
  \textbf{06} (2013) 065},
  \href{http://arxiv.org/abs/1304.6865}{{\normalfont\ttfamily
  arXiv:1304.6865}}\relax
\mciteBstWouldAddEndPuncttrue
\mciteSetBstMidEndSepPunct{\mcitedefaultmidpunct}
{\mcitedefaultendpunct}{\mcitedefaultseppunct}\relax
\EndOfBibitem
\bibitem{Aaij:2013cby}
LHCb collaboration, R.~Aaij {\em et~al.},
  \ifthenelse{\boolean{articletitles}}{\emph{{Search for the lepton-flavor
  violating decays $B^0_s \rightarrow e^{\pm}\mu^{\mp}$ and $B^0 \rightarrow
  e^{\pm} \mu^{\mp}$}},
  }{}\href{http://dx.doi.org/10.1103/PhysRevLett.111.141801}{Phys.\ Rev.\
  Lett.\  \textbf{111} (2013) 141801},
  \href{http://arxiv.org/abs/1307.4889}{{\normalfont\ttfamily
  arXiv:1307.4889}}\relax
\mciteBstWouldAddEndPuncttrue
\mciteSetBstMidEndSepPunct{\mcitedefaultmidpunct}
{\mcitedefaultendpunct}{\mcitedefaultseppunct}\relax
\EndOfBibitem
\bibitem{ATLAS-CONF-2010-009}
ATLAS collaboration, \ifthenelse{\boolean{articletitles}}{\emph{{Estimating
  Track Momentum Resolution in Minimum Bias Events using Simulation and $K_{\rm
  S}$ in $\sqrt{s} = 900$ GeV collision data}}, }{} Tech. Rep.
  ATLAS-CONF-2010-009, CERN, Geneva, Jun, 2010\relax
\mciteBstWouldAddEndPuncttrue
\mciteSetBstMidEndSepPunct{\mcitedefaultmidpunct}
{\mcitedefaultendpunct}{\mcitedefaultseppunct}\relax
\EndOfBibitem
\bibitem{Aaij:2012nna}
LHCb collaboration, R.~Aaij {\em et~al.},
  \ifthenelse{\boolean{articletitles}}{\emph{{First Evidence for the Decay
  $B_s^0 \to \mu^+ \mu^-$}},
  }{}\href{http://dx.doi.org/10.1103/PhysRevLett.110.021801}{Phys.\ Rev.\
  Lett.\  \textbf{110} (2013), no.~2 021801},
  \href{http://arxiv.org/abs/1211.2674}{{\normalfont\ttfamily
  arXiv:1211.2674}}\relax
\mciteBstWouldAddEndPuncttrue
\mciteSetBstMidEndSepPunct{\mcitedefaultmidpunct}
{\mcitedefaultendpunct}{\mcitedefaultseppunct}\relax
\EndOfBibitem
\bibitem{Aaboud:2016ire}
ATLAS collaboration, M.~Aaboud {\em et~al.},
  \ifthenelse{\boolean{articletitles}}{\emph{{Study of the rare decays of
  $B^0_s$ and $B^0$ into muon pairs from data collected during the LHC Run 1
  with the ATLAS detector}},
  }{}\href{http://dx.doi.org/10.1140/epjc/s10052-016-4338-8}{Eur.\ Phys.\ J.\
  \textbf{C76} (2016), no.~9 513},
  \href{http://arxiv.org/abs/1604.04263}{{\normalfont\ttfamily
  arXiv:1604.04263}}\relax
\mciteBstWouldAddEndPuncttrue
\mciteSetBstMidEndSepPunct{\mcitedefaultmidpunct}
{\mcitedefaultendpunct}{\mcitedefaultseppunct}\relax
\EndOfBibitem
\end{mcitethebibliography}

\end{document}